\documentclass[preprint]{aastex}

\newcommand{\magn}{\(\stackrel{^m}{_.}\)}

\shortauthors{Parodi et al.}
\shorttitle{Supernova Type Ia Luminosities}

\begin{document}

\title{\sc Supernova Type Ia Luminosities, Their Dependence on Second
Parameters, and the Value of H$_0$}

\author{\sc B. R.~Parodi\altaffilmark{1}, A.~Saha\altaffilmark{2},
A.~Sandage\altaffilmark{3}, and G. A.~Tammann\altaffilmark{1}}

\altaffiltext{1}{Astronomisches Institut der Universit\"at Basel,
Venusstr.~7, CH-4102 Binningen, Switzerland; parodi@astro.unibas.ch.}
\altaffiltext{2}{National Optical Astronomy Observatories, 950 North
Cherry Avenue, Tucson, AZ 85726-6732, USA; saha@noao.edu.}
\altaffiltext{3}{The Observatories of the Carnegie Institution of 
Washington, 813 Santa Barbara Street, Pasadena, CA 91101-1292, USA.}

%------------------------------------------------------------------
\begin{abstract}
A sample of 35 SNe\,Ia with good to excellent photometry in B and V,
minimum internal absorption, and 1200 $<$ $v \stackrel{_<}{_\sim}$
30\,000\,km\,s$^{-1}$ is compiled from the literature. As far as their
spectra are known they are all Branch-normal. For 29 of the SNe\,Ia
also peak magnitudes in I are known. The SNe\,Ia have uniform colors
at maximum, i.e. $<$$B$-$V$$>$=$-$0\magn012 ($\sigma$=0.051) and
$<$$V$-$I$$>$=$-$0\magn276 ($\sigma$=0.078). In the Hubble diagram
they define a Hubble line with a scatter of
$\sigma_M$=0\magn21-0\magn16, decreasing with wavelength. The scatter
is further reduced if the SNe\,Ia are corrected for differences in
decline rate $\Delta m_{15}$ or color ($B$$-$$V$). A combined
correction reduces the scatter to
$\sigma$\,$\stackrel{_<}{_\sim}$\,0\magn13. After the correction no
significant dependence remains on Hubble type or galactocentric
distance. The Hubble line suggests some curvature which can be
differently interpreted. A consistent solution is obtained for a
cosmological model with $\Omega_M$=0.3, $\Omega_\Lambda$=0.7, which is
indicated also by much more distant SNe\,Ia. Absolute magnitudes are
available for eight equally blue (Branch-normal) SNe\,Ia in spirals,
whose Cepheid distances are known. If their well defined mean values
of $M_B$, $M_V$, and $M_I$ are used to fit the Hubble line to the
above sample of SNe\,Ia one obtains $H_0$=58.3
km\,s$^{-1}$\,Mpc$^{-1}$, or, after adjusting all SNe\,Ia to the
average values of $\Delta m_{15}$ and ($B$$-$$V$), $H_0$=60.9
km\,s$^{-1}$\,Mpc$^{-1}$. Various systematic errors are discussed
whose elimination tends to decrease $H_0$. The finally adopted value
at the 90-percent level, including random and systematic errors, is
$H_0$=58.5$\pm$6.3 km\,s$^{-1}$\,Mpc$^{-1}$. Several higher values of
$H_0$ from SNe\,Ia, as suggested in the literature, are found to
depend on large corrections for variations of the light curve
parameter and/or on an unwarranted reduction of the Cepheid distances
of the calibrating SNe\,Ia.
\end{abstract}

 
%------------------------------------------------------------------
\section{Introduction}

Supernovae of type Ia (SNe Ia) are the prime distance indicators for
the determination of the Hubble constant $H_0$ since they can be
followed out to large distances, and since it is possible to determine
Cepheid distances with HST for the nearest of their host galaxies and
hence to calibrate their luminosity at maximum \citep{sand92}. This
immediately yields the distances of the more distant SNe Ia because
they are -- if restricted to blue objects -- the best standard candles
known with a luminosity scatter of less than
$\sigma_M$\,=\,0\magn25. As standard candles they are now so heavily
relied upon that they are even used for the much more sensitive test
for the cosmological constant $\Lambda$ \citep{ries98b,perl99}. A
wealth of excellent photometric data for SNe Ia has been accumulated
since 1985 by the Cal\'an/Tololo team \citep{hamu96b} and others. They
reach out to \( \sim30\,000\) km\,s$^{-1}$, i.e. far enough to avoid
the effects of peculiar and streaming motions. In parallel, the HST
Supernova Project has by now provided Cepheid-calibrated luminosities
for six nearby SNe Ia \citep{saha99}, not counting SN 1895 B for which
only the B magnitude is known. These are augmented by two additional
calibrators from \citet{tanv95} and \citet{turn98}. With a total of
eight calibrators and three dozen more distant SNe Ia, the statistics
rests on solid ground.

The one remaining question is whether the calibrators and the distant
SNe Ia are genuine twins, or whether the different selection criteria
cause systematic differences between the samples. For instance, the
Cepheid distances of the calibrators imply that they lie in galaxies
containing young-population stars, whereas the distant SNe Ia have
also been observed in S0 and E galaxies. Various second parameters
have been proposed to correlate with SN luminosity; they can serve as
a control of luminosity differences between the nearby calibrators and
the distant SNe Ia. From the point of view of the physicist the most
interesting second parameters are several spectral features which are 
known to correlate with luminosity
\citep{nuge95,fish95,fish99,ries98a,mazz98}. But the available data
are too sparse to be useful in the present context. One is therefore
reduced to empirical second parameters such as light curve shape, SN
color, Hubble type and position in the parent galaxy. Second
parameters have been discussed by, e.g.,
\citet{phil93,tamm95,vaug95,hamu96a,hamu96c,trip98,trip99,saha99,ries99,jha99,phil99,sunt99,gibs00a}.

The difficulty to find a correlation between SN Ia luminosities and
the second parameters in face of an intrinsic scatter of
$\sigma_M$$\stackrel{_<}{_\sim}$0\magn25 is that very accurate
{\em relative} distances are required. Cepheid distances and, e.g.,
Tully-Fisher distances are not sufficiently accurate for the
purpose. SNe Ia in the Virgo cluster cannot be assumed to lie at the
same distance because of the important depth effect of the
cluster. Regress must therefore be taken to the {\em relative} distances
that are indicated by recession velocities. All velocities $v$\,$>$\,1200
km\,s$^{-1}$ are taken as indicative of the relative distances. This
is permissible because the errors assigned to the relative distances
and to the resulting relative absolute magnitudes make allowance for
reasonable values of the peculiar velocities (Section 2.1).

The purpose of the present paper is to discuss the correlation of
second parameters with the peak luminosity of SNe\,Ia --- using an
enlarged and well-defined sample of distant SNe Ia --- and to
determine the value of the Hubble constant. The SNe Ia magnitudes,
after correction for decline rate $\Delta m_{15}$ and color ($B$-$V$),
have a scatter of only $\sigma_M$$\stackrel{_<}{_\sim}$0\magn13. In
fact the magnitude-corrected SNe Ia define the Hubble diagram so well
that a flat Universe model with $\Omega_M$=0.3, $\Omega_\Lambda$=0.7
gives a marginally better fit than an $\Omega_M$=1 model. The corrected
magnitudes, if combined with the corrected absolute magnitudes of eight
Cepheid-calibrated SNe\,Ia, determine $H_0$ with a very small {\em
statistical} error.

The organization of the present paper is as follows. In Section 2 the
available data for blue SNe Ia are compiled, and their luminosity
calibration by means of eight SNe Ia with known Cepheid distances is
discussed.   The SNe Ia colors and extinctions are discussed in
Section 3. The Hubble diagram is shown in Section 4. In Section 5 the
SN Ia luminosities are discussed in function of decline rate $\Delta
m_{15}$, SN color ($B$-$V$), Hubble type $T$, and position in the
parent galaxy. The effective Hubble diagram with decline-rate and
color corrected magnitudes is shown in Section 6, and the resulting
values of $H_0$ are derived. Alternative solutions are explored in
Section 7, and the conclusions are given in Section 8.

%------------------------------------------------------------------
\section{The Photometric Data}

\subsection{Blue SNe\,Ia within v $\stackrel{_<}{_\sim}$ 30\,000 km\,s$^{-1}$}

The available data for all SNe Ia (n=67) with
($B_{max}$-$V_{max}$)$\le$ 0.20\footnote{In the following we write for
($B_{max}$-$V_{max}$) and ($V_{max}$-$I_{max}$) more conveniently
(B-V) and (V-I).} (after correction for Galactic reddening following
\citet{schl98}), and with
$v$\,$\stackrel{_<}{_\sim}$\,30\,000\,km\,s$^{-1}$ (the two largest
accepted velocities are $v$\,=\,30\,269\,km\,s$^{-1}$ of SN\,1992\,aq
and $v$\,=\,37\,325\,km\,s$^{-1}$ of SN 1996 ab) are compiled in Table
1.

\placetable{tbl1}
 
The individual columns bear the following informations:\\ {\bf(1)}:
The supernova designation. If followed by an acceptance sign $\surd$
the SN is included in the fiducial sample (see below). {\bf(2)}: The
Hubble type of the parent galaxy as coded by \citet{deva74}, but
slightly simplified for the early and latest types: E: T=-3; E/S0:
T=-2; S0: T=-1; S0/a: T=0; Sa: T=1; Sab: T=2; Sb: T=3; Sbc: T=4;
Sc,Sd,Sm \& Im: T=5. The Am galaxy NGC 5253 (SN 1972 E) has
tentatively been ascribed the type T=5. {\bf(3)}: The decimal
logarithm of the galaxy redshift velocity cz. Most redshifts are from
the Lyon/Meudon Extragalactic Database (LEDA;
http://www-obs.univ-lyon1.fr); additional redshifts were taken from
\citet{hamu96b} and \citet{ries99}. They were corrected for the motion
of the Sun relative to the centroid of the local group \citep{yahi77}
and for a self-consistent Virgocentric infall model with a local
infall vector of 220\,km\,s$^{-1}$ \citep{kraa86}; beyond $v_{220}=
3000$\,km\,s$^{-1}$ an additional correction for the motion of
630\,km\,s$^{-1}$ relative to the CMB dipole anisotopy \citep{smoo92}
was applied. Varying the size of the local co-moving volume between
2000 and 10\,000 km\,s$^{-1}$ has no significant effect on the present
conclusions. For the members of three clusters the following mean
velocities were assumed: for Cen A group members
$v_{220}\,=\,291$\,km\,s$^{-1}$ \citep{kraa86}, for Virgo cluster
members, as assigned by \citet{bing93},
$v_{220}\,=\,1179$\,km\,s$^{-1}$ \citep{jerj93}, for Fornax cluster
members, as assigned by \citet{ferg88},
$v_{220}\,=\,1440$\,km\,s$^{-1}$ \citep{tamm97}. The errors of $\log
v$ in units of 0.01 are shown in parentheses; they are compounded from
the observational errors and the peculiar velocities, assumed to be
200 km\,s$^{-1}$ within 1500 km\,s$^{-1}$, 400 km\,s$^{-1}$ for 1500
$<$ v $<$ 3000 km\,s$^{-1}$, and 600 km\,s$^{-1}$ beyond 3000
km\,s$^{-1}$. {\bf(4)-(6)}: $B$-, $V$-, and $I$-band apparent peak
magnitudes. They are in the Cerro Tololo system of fitting light curve
templates \citep{hamu96b,hamu96c}. The template fitting for the
observations by \citet{ries99} was done by us. All magnitudes are
corrected for Galactic absorption. Where applicable, the K-correction
\citep{hamu93} was taken into account. The magnitude errors in units
of 0\magn01 were taken from the original literature or were estimated
by us. {\bf(7)}: The Galactic absorption $A_V$ from \citet{schl98},
assuming $A_{B}$=4.3\,$E$($B$-$V$), $A_{V}$=3.3\,$E$($B$-$V$), and
$A_{I}$=2.0\,$E$($B$-$V$) throughout. {\bf(8)}: $\Delta m_{15}$ being
the decline in magnitudes of the $B$ light curve during the first 15
days after maximum, as defined by \citet{phil93}. The magnitude errors
in units of 0\magn01 were taken from the original literature or were
estimated by us. {\bf(9)}: The onset of the photometric data given in
days before (minus signs) and after (plus signs) $B$
maximum. {\bf(10)}: References for the photometric data $B$, $V$, $I$
and $\Delta m_{15}$. {\bf(11)-(13)}: Absolute magnitudes at maximum as
calculated from the apparent magnitudes in columns (4) to (6) and from
the recession velocities in column (3). A flat matter universe
($\Omega_M$=1) and a value of $H_{0}=60$\,km\,s$^{-1}$\,Mpc$^{-1}$ is
assumed. As a consequence all listed absolute magnitudes scale, except
for peculiar velocities, with $5\,\log$($H_{0}/60$). The errors in
units of 0\magn01 are compounded from the errors given in columns (3)
and (4) to (6), respectively. No absolute magnitudes were calculated
for SNe\,Ia in the field with $v_{220}\,<\,1200$\,km\,s$^{-1}$ because
their velocity distances are too unreliable due to peculiar motions,
and for five SNe Ia in Virgo cluster galaxies because of the
considerable cluster depth. For four SNe Ia of Table\,1 independent
absolute magnitudes from Cepheid distances are given in Table 3; they
are marked with asterisks.

In the following the overluminous SN 1991\,T \citep{phil92} and its
twin SN 1995\,ac \citep{garn96} are left out because they are
spectroscopically peculiar, leaving in Table\,1 59 blue SNe with
($B$-$V$)$\le$0.20 and $v_{220}\ge \,1200$\, km\,s$^{-1}$.  The
observations of six SNe Ia after 1985 in Table\,1 begin only eight
days after maximum or later. Their extrapolated maximum magnitudes may
be less accurate \citep{phil99}. In the diagrams to follow they are
shown with small symbols, but they have no systematic effect on any of
the conclusions below and are included in the equations below with
their errors as given in Table\,1. The I-observations of SN 1992\,au
begin only 15 days after $B$ maximum and are not considered in the
following.

\placetable{tbl2} 

In Table 2 data are compiled that are relevant for a localization of
the SNe Ia of Table\,1 within their host galaxies. Columns (1) to (6)
are self-explanatory. The diameters $D_{25}$ (in arcsec), given in
column (7) where available, are taken from the on-line Asiago
Supernova Catalogue \citep{capp98} and the RC3 \citep{deva91}. Columns
(8) and (9) give the supernova offsets in the E/W and N/S directions
as taken from \citet{capp98} and from \citet{ries99}. The projected
galactocentric distances of the SNe in units of the galaxy radius
$r_{25}$ (=$D_{25}/2$) are given in column (10); they are
distance-independent.

\subsection{The Cepheid-calibrated SNe\,Ia}

Table\,3 lists the nine SNe Ia for which Cepheid distances and
therefore absolute magnitudes are known. The apparent or true distance
moduli from Cepheids and their sources are given in columns (4) to
(7). The moduli from the WFPC 2 are corrected by +0\magn05 for the
photometric short exposure/long exposure zeropoint effect
\citep{stet95,saha96a}. The apparent $B$, $V$, and $I$ magnitudes at
maximum (uncorrected for Galactic absorption) and their sources are in
columns (8) to (11). The total color excesses E$_{B-V}$ (Galactic and
within the host galaxy) and their sources are listed in columns (12)
to (13). The absorption-corrected apparent magnitudes B$^0$, V$^0$,
and I$^0$ are given in columns (14) to (16). The absolute magnitudes
$M_B^0$, $M_V^0$, and $M_I^0$ follow in columns (17) to (19). For the
first four SNe Ia the absolute magnitudes are derived by subtracting
the {\em apparent} distance moduli from the respective {\em apparent}
magnitudes on the plausible assumption that the Cepheids and the
corresponding SN Ia suffer the same (small) amount of absorption. In
the remaining cases either the Cepheids have variable absorption or
the SN suffers additional absorption in its host galaxy. In these
cases the absolute magnitudes come from subtracting the {\em true}
distance moduli from the {\em true} magnitudes. Finally, the decline
rates $\Delta m_{15}$ in column (20) are taken from the references in
column (11). The straight and weighted mean absolute magnitudes of
eight SNe Ia in Table 3 are given at the bottom of the Table. The
bright SN 1895\,B is not included because its V maximum is known too
poorly. --- The adopted absolute magnitudes are in fortuitous
agreement with theoretical models of {\em blue} supernovae
\citep{hoef96} and Branch's (1998) discussion of the physical
properties of SNe\,Ia.

\placetable{tbl3} 

Five SNe in Table 3 are not included in Table 1. SN 1895\,B has no
reliable color information. SNe 1989\,B and 1998\,bu are observed to
be much redder than (B-V)=0.20. SNe 1974\,G and 1981\,B are {\em
known}, in spite of their rather blue color, to suffer reddening in
their host galaxy.

The eight Cepheid distances adopted in Table 3 have been re-analysed
by \citet{gibs00} (hereafter G00). They have used ALLFRAME photometry
and the associated method of detecting variable stars. They claim that
this re-analysis places their galaxy distances on the same footing as
the distances of galaxies contained in the Mould-Freedman-Kennicutt
(MFK; e.g. Mould et al. 2000) program. However, this claim is questionable
for several reasons:

(i) While the MFK team has consistently quoted photometry from
ALLFRAME (though DoPHOT magnitudes have also been presented for
comparison), their list of Cepheids comprised objects independently
detected by both ALLFRAME {\em and} DoPHOT. The Sandage/Saha
consortium have based their Cepheid selection and photometry on DoPHOT
alone. A salient result of the G00 re-analysis is that the agreement
of the photometry of stars that are {\em in common} between them and
the Sandage/Saha team must be excellent, because if one compares only
the 118 Cepheids in common in six galaxies (as given in Table\,3 of
G00), the difference in the magnitudes averages to be $\Delta
V$=-0\magn044$\pm$0\magn002, $\Delta I$=-0\magn038$\pm$0\magn004,
in the sense that the ALLFRAME-based magnitudes are brighter. Thus if
the procedure that was actually used to derive distances by the MFK
team in their program is applied, the agreement between ALLFRAME- and
DoPHOT-based magnitudes in $V$ and $I$, and hence in moduli, is as good
as one can expect, and the confidence in the DoPHOT-based results is
bolstered.

However, G00 have added Cepheids found only by ALLFRAME. These
``extra'' Cepheids drive a distinctly different result, increasing the
mean difference DoPHOT-moduli (of the Sandage/Saha team) minus
ALLFRAME-moduli (of G00) to 0\magn17. Had G00 reported results that
are truly on the same footing as the rest of the MFK team analysis,
they would not have seen the 0\magn17 ``discrepancy''.

ii) G00 specially point out that in the case of NGC 4536 Saha's et
al. (1996a) Cepheids in Chip 2 of the WFPC2 give  a distance modulus
that is larger by 0.66 mag, which is the result of  differences in the
photometry in V and/or I by order of 0.25 mag. G00 do not see this
discrepancy in their own reduction: their photometry in all four chips
are  in mutual agreement, and also in nominal agreement  with Saha's
et al. (1996a) photometry in chips 1, 3, and 4. It is worth emphazising
that the region of the galaxy imaged in  chip 2 is different in
character from that  in chips 3 and 4 (chip 1 is similar to chip 2,
but there are too few Cepheids to make a statistically significant
difference):  the latter are dominated by the outer spiral arms, but
chip 2 shows the more amorphous inner regions. Such changes in
environment can contribute different levels of  confusion noise, and
consequently result in mis-calibration of the zero-point.  To be
consistent with their work in other galaxies and having no external
information on which chip is best, Saha et al. (1996a) have combined
their Cepheids, as measured, in a single PL relation and have derived
an average distance modulus, their rationale being that the
exceptional chip-to-chip variation of NGC 4536 may just be the result
of an unusually large {\em statistical} fluctuation.  The point
illustrates that the chip-to-chip variations of the aperture
correction in crowded regions is the weakest link in WFPC2
photometry. Cepheid moduli from the WFPC2 therefore always carry an
uncertainty of 0.10-0.15 mag. The error is  random from galaxy to
galaxy and is therefore  reduced by a sufficient number of
calibrators.

(iii) The analysis of G00 differs also from the present one in the way
the reddening is handled, both of the Cepheids as well as of the SNe
Ia. The G00 approach is to always de-redden the Cepheids, and then to
obtain the de-reddened modulus to the host galaxy. This approach is
unavoidable if the extinction of the Cepheids is much larger than that
of the SN\,Ia (NGC\,4639 with SN\,1990\,N), if the Cepheids have
differential reddening (NGC\,3627), or/and if the SNe\,Ia have large,
but pre-determinded reddenings (SN\,1974\,G, 1981\,B, 1989\,B, and
1998\,bu). Yet de-reddening procedures demand exquisite photometry,
since the ratio of total to selective absorption amplifies the
uncertainties in the photometry. Therefore, instead of going through
the de-reddening twice (once for the Cepheids and once for the
SNe\,Ia), it is preferable to assume that the reddening of the
Cepheids and of the SNe\,Ia is the same whenever the reddening is
small ($\stackrel{_<}{_\sim}$0\magn03). In that case the SN luminosity
is obtained by simply subtracting the apparent distance modulus from
the apparent SN magnitude. This procedure has been applied for the
first four entries in Table\,3.

After the reddening corrections of G00 are applied, the
``discrepancy'' of 0\magn17 drops to 0\magn04, 0\magn07, and 0\magn15,
in $M_B$, $M_V$, and $M_I$, respectively. The consequence of this
calibration then is --- if applied to the fiducial sample --- that the
value of $H_0(I)$ becomes larger by five percent than $H_0(B)$ while
with the present calibration in Table\,3 the values of $H_0$ agree
almost exactly in all three passbands  (cf. Section 4).

Remaining systematic error sources, which may affect the calibrators,
are discussed in Section\,6.

%------------------------------------------------------------------
\section{The Colors of SNe Ia}

\subsection{Justification for the {\em blue} sample of SNe Ia}

In Table\,1 and in the following sections only SNe Ia with
($B$-$V$)$\le$0\magn20, called blue SNe\,Ia, are considered. This
color cut needs justification. In Figure\,1 {\em all} SNe Ia between
1985 and 1996 with known peak magnitudes are plotted, unrestricted as to 
color ($B$-$V$). The restriction to SNe\,Ia after 1985 is here and in
the following indicated because the advent of CCD photometry has much
improved the photometric accuracy. The SNe\,Ia colors are only
corrected for Galactic reddening. 44 SNe Ia are strongly concentrated
toward a mean color of ($B$-$V$)=0\magn02 with a scatter of only
$\sigma_{B-V}$=0\magn05. This small scatter is even the more
surprising as Hamuy et al. (1996b, Table 6) list an average
observational error in color of 0\magn06. The {\em true} color scatter
of this subsample could therefore be arbitrarily small. As far as
spectra of this blue subsample are known they are all Branch-normal
\citep{bran93}, the only exceptions being the twins
SN 1991\,T and 1995\,ac  with peculiar early specta. As mentioned
before they are excluded in the following.

\placefigure{fig1} %BmV_Distribution

The 10 SNe\,Ia in Fig.\,1 with ($B$-$V$)$>$0\magn20 contain a
high fraction of spectroscopically peculiar objects, like the very red
and strongly underluminous SN 1991\,bg and the low-expansion velocity
SN 1986\,G and their counterparts, as well as some presumably strongly
reddened SNe\,Ia. These red SNe\,Ia clearly form a very heterogeneous
group.

It is obvious that if SNe\,Ia are to be used as standard candles, one
must concentrate on the homogeneous class of {\em blue} SNe\,Ia.

\subsection{The true colors of SNe\,Ia}

Even the sample of 42 blue SNe\,Ia may be affected by some internal
reddening. Their true colors are best reflected by SNe Ia that have
occured in E/S0 galaxies or lie in the outer regions of
spirals. Outlying SNe\,Ia are here defined as having $r/r_{25}$$>$0.4,
where $r$ is the radial distance from the center of the host galaxy
(in arcsec), and $r_{25}$ is the de Vaucouleurs radius of that galaxy
(in arcsec). The $r/r_{25}$ values are listed in Table 4 for all SNe
for which the necessary data are available. The relative radial
distance $r/r_{25}$$>$0.4 corresponds roughly to the limit to which the
spiral structure can be traced.

\placetable{tbl4}

Table 4 lists the mean colors $<$$B$-$V$$>$ and $<$$V$-$I$$>$ of all
SNe\,Ia in E/S0 galaxies, of the outlying SNe\,Ia in spirals, and of
the calibrators in Table 2. The three groups have closely the same
mean colors, i.e. $<$$B$-$V$$>$=-0.012$\pm$0.051 and
$<$$V$-$I$$>$=-0.276$\pm$0.078. The overall means are therefore
adopted as the true colors of SNe\,Ia in E/S0s and spirals alike.

\citet{phil99} have instead derived the intrinsic colors by
assuming that all SNe\,Ia with $E$($B$-$V$)$_{Tail}$$<$0.06 are
unreddened. (For their new method to derive $E$($B$-$V$)$_{Tail}$ see
the original paper). They give $E$($B$-$V$) and $E$($V$-$I$) values
for 40 SNe\,Ia with B- and for 32 SNe\,Ia with I-magnitudes in their
Table 1. If these color excesses are applied to the magnitudes in
Table 1 one obtains mean colors of $<$$B$-$V$$>$=-0.051$\pm$0.007 and
$<$$V$-$I$$>$=-0.331$\pm$0.013, i.e. noticeably bluer by
$\sim$0\magn04 than adopted in Table 4. It seems therefore that the
color excesses of \citet{phil99} are too large. The difference
is not trivial when converted to absorption, i.e. $A_B$$\approx$0.16;
it has, however, no effect on $H_0$ as long as the adopted colors of
the calibrators and the distant SNe Ia are the same.

\subsection{Blue, yet reddened SNe\,Ia}

The question remains whether some of the SNe\,Ia in the inner regions
of the spirals, i.e. $r/r_{25}<0.4$, are affected by internal
absorption. Figure 2 shows the correlation as to color ($B$-$V$) with radial
distance for all SNe\,Ia for which the necessary data are
available. There are five inner SNe\,Ia in spirals which are redder than
($B$-$V$)=0\magn06. Since their absolute magnitudes are also fainter than
average they are strong candidates for some internal absorption. They are
listed in Table 5 together with two additional SNe\,Ia that fullfill
the same color and magnitude requirement, but for which $r/r_{25}$ is
not available.

\placefigure{fig2}
\placetable{tbl5}

The colors ($B$-$V$) and ($V$-$I$) in columns 3 and 4 of Table 5 are
from data in Table\,1. These colors and the adopted mean color in
Table\,4 give the excesses $E$($B$-$V$) and $E$($V$-$I$). If these are
averaged, giving double weight to the former and assuming
$E$($B$-$V$)=0.6\,$E$($V$-$I$), one obtains the mean excesses
$E$($B$-$V$) in column 5, which in turn lead to the
absorption-corrected apparent (columns 6 to 8) and absolute (columns 9
to 11) magnitudes of the SNe\,Ia. They are on average close to the
absolute magnitudes of the unreddened SNe\,Ia. \citet{phil99} have
given reddening values for four objects in Table 5; they are also
above average. The strongest argument for the internal extinction is,
however, that at given $\Delta m_{15}$ they are fainter on average by
0\magn38 in $B$, 0\magn30 in $V$, and 0\magn14 in $I$ than their
unreddened counterparts (cf. below, Section 5.1). This magnitude
difference practically disappears once the extinction corrections have
been applied. Because the seven SNe Ia of Table 5 being reddened
cannot be proved beyond doubt, they are excluded henceforth.  If they
had been retained without absorption correction they would slightly
decrease the value of $H_0$; if included after the absorption
correction they would not have a net effect on $H_0$.

It is reassuring that the remaining 10 SNe Ia in spirals, which have
$r/r_{25}$$<$0.4 or for which no information is available as to radial
distance, have closely the same mean colors as the unreddened SNe\,Ia
(Table 4).

%------------------------------------------------------------------
\section{The Hubble Diagram}

After the exclusion of SNe 1991\,T and 1995\,ac, and the I-band
magnitude of SN 1992\,au, as well as the seven SNe\,Ia in Table\,5,
Table\,1 contains 35 (29 of which have also I-magnitudes) blue,
unreddened SNe Ia after 1985 and with v$>$1200\,km\,s$^{-1}$. They are
referred to in the following as the {\em fiducial sample}. Their
Hubble diagrams in $B$, $V$, and $I$ are shown in Fig.\,3. Fitted to the
data is a Hubble line assuming a flat Universe with $\Omega_M$=0.3 and
$\Omega_\Lambda$=0.7 (for a justification of this choice cf. Section
6). In that case the Hubble line is described by (cf. Carroll, Press,
\& Turner 1992)
\begin{equation}
m_{B,V,I}=5\log \left( \frac{c}{H_0} (1+z_1)\int_{0}^{z_1} [(1+z)^2(1+\Omega_Mz)-z(2+z)\Omega_\Lambda]^{-1/2}dz \right) + M_{B,V,I}+25\,.
\end{equation}
Inserting the weighted values M$_{B,V,I}$ of the calibrators from
Table\,3 the best fit to the data is achieved by weighted $\chi^2$
solutions in B, V, and I. They give as a preliminary result a
Hubble constant\footnote{In the remainder of this paper a value of the
Hubble constant is always understood in terms of
km\,s$^{-1}$\,Mpc$^{-1}$.} of $H_0$(B)=58.3$\pm$1.1,
$H_0$(V)=57.9$\pm$1.8, and $H_0$(I)=58.8$\pm$2.6, with a mean value of
\begin{eqnarray} 
  H_0(BVI)&=&58.3\pm2.0\,. 
\end{eqnarray}
The scatter about the Hubble lines in Fig.\,3 is $\sigma_B$=0.214,
$\sigma_V$=0.181, and $\sigma_I$=0.161 mag, proving in favor of the use of
SNe Ia as standard candles. The scatter is somewhat larger within
$v$=10\,000\,km\,s$^{-1}$ than beyond, which must be due to the
influence of peculiar motions. The quoted values of $\sigma$ are
therefore upper limits of the intrinsic luminosity scatter.

\placefigure{fig3}

%------------------------------------------------------------------
\section{The Correlation of SN Ia Luminosities with Second Parameters}

Even the small scatter of $\sigma _B$=0\magn21 in the Hubble diagram of
the fiducial sample spans a total range of SN Ia luminosities of
$\sim$0\magn6. This is enough of a variation to ask whether their
individual luminosities depend on second parameters. If that is the
case one must make sure that the calibrators in Table\,3 have the same
mean second parameters as the SNe of the fiducial sample.

Even the SNe Ia with Branch-normal spectra \citep{bran93} show some
spectral variations which apparently correlate with the expansion
velocity, the effective temperature --- presumably a result of
variable amounts of $^{56}$Ni produced in the explosion ---, {\em and}
the peak luminosity (\citet{nuge95}, cf. also
\citet{mazz98}). Attempts to homogenize all blue SNe Ia to a common
mean spectrum and thus to make them even better standard candles are
doomed, as stated before,  because too few SNe\,Ia of the fiducial
sample have the necessary spectral information.

As a consequence one has searched for purely empirical second
parameters that correlate with the peak luminosity. The first success
was the decline rate $\Delta m_{15}$ of the $B$ light curve, which
measures the decline in magnitudes during the first 15 days after $B$
maximum \citep{phil93}. Other second parameters followed like the SN
light curve shape \citep{ries96,perl98}, the SN color ($B$-$V$) at $B$
maximum \citep{tamm82,trip98}, the color \citep{bran96b} or the
Hubble type \citep{hamu95,saha97,saha99} of the parent galaxy, or the
position within the latter \citep{wang97,ries99}. There is a
considerable literature on the subject.

The dependence of SNe\,Ia luminosities on second parameters is
re-investigated here on the basis of an enlarged, well-defined sample.

\subsection{Decline Rate $\Delta m_{15}$}

The B, V, I residuals, read in magnitudes, from the Hubble line in
Fig.\,3, i.e. $\delta m$\,=\,$m_{obs}-m_{fit}$, are plotted
versus the decline rate $\Delta m_{15}$ in Fig.\,4.

\placefigure{fig4}

The clear dependence of SN luminosity on $\Delta m_{15}$ is expressed
by the following linear regressions (weighted by the errors in $M_{B,V,I}$ as
given in Table\,1, columns (11) to (13)\,):
\begin{eqnarray}
  \delta m_B^{15}&=&0.48_{\pm0.13}\,(\Delta m_{15}-1.2)-28.410_{\pm0.161}, \;\;\;\sigma=0.194,\;\;\;n=35\\
  \delta m_V^{15}&=&0.50_{\pm0.11}\,(\Delta m_{15}-1.2)-28.394_{\pm0.143}, \;\;\;\sigma=0.151,\;\;\;n=35\\
  \delta m_I^{15}&=&0.39_{\pm0.13}\,(\Delta m_{15}-1.2)-28.118_{\pm0.171}, \;\;\;\sigma=0.144,\;\;\;n=29
\end{eqnarray}
If the linear regressions are made by allowing for the errors both in
$m$ {\em and} $\Delta m_{15}$ one finds somewhat steeper
slopes. However, any steeper slope introduces a dependence of the
luminosity on $\Delta m_{15}$ of opposite sign (i.e. SNe Ia with large
values of $\Delta m_{15}$ become too bright). If errors both in $m$
and $\Delta m_{15}$ are considered one has to impose the additional
condition that any dependence of absolute magnitude (or $\delta m$) on
$\Delta m_{15}$ vanishes. In this case one recovers equations (3) to
(5). Some authors have also  suggested steeper slopes by forcing a
relation through all kinds of SNe\,Ia including the heterogenous set
of red objects (cf. Section\,3.1). Equations (3) to (5) apply
explicitely only to SNe\,Ia which fullfill the conditions imposed on
the fiducial sample. The decisive point is that these conditions (with
the exception of recession velocity), i.e. reasonably good photometry
in $B$ and $V$, blue intrinsic colors ($B$-$V$), and Branch-normal
spectra whereever known, are perfectly met also by the calibrators,
and hence they must comply with the same equations. (It may be noted
in passing that the calibrators alone suggest an even somewhat
shallower slope).

Correcting the SNe Ia magnitudes for $\Delta m_{15}$ according to
equations (3) to (5) has two effects: i) it reduces the scatter in $m$
to the indicated values, and ii) it increases $H_0$ because the
calibrators have somewhat smaller $\Delta m_{15}$ on average than the
fiducial sample (cf. Section 6 below).

\subsection{SN color ($B$-$V$)}

The rediduals $\delta m$\,=\,$m_{obs}-m_{fit}$ from the Hubble line in
Fig.\,3 are plotted versus SN color ($B$-$V$) in Fig.\,5. Weighted
linear regressions through the data in Fig.\,5 give
\begin{eqnarray}
  \delta m_B^{\rm col}&=&2.55_{\pm0.46}\,[(B-V)+0.01]-28.371_{\pm0.027}, \;\;\;\sigma=0.160,\;\;\;n=35\\
  \delta m_V^{\rm col}&=&1.50_{\pm0.40}\,[(B-V)+0.01]-28.348_{\pm0.024}, \;\;\;\sigma=0.164,\;\;\;n=35\\
  \delta m_I^{\rm col}&=&1.17_{\pm0.43}\,[(B-V)+0.01]-28.083_{\pm0.029}, \;\;\;\sigma=0.148,\;\;\;n=29
\end{eqnarray}

In spite of the considerable observational errors of ($B$-$V$) the
dependence on luminosity is significant. In fact comparison of the
luminosity scatter in equations (6) to (8) with the ones in equations
(3) to (5) shows that ($B$-$V$) is as efficient as the decline rate
$\Delta m_{15}$ to reduce the scatter. This is in line with the
conclusion of \citet{trip98}. The dependence of the SN Ia luminosity
on color must be mainly an {\em intrinsic} effect of SNe Ia
\citep{tamm95}. The proposal that it is due to absorption in the
parent galaxy \citep{ries96} has been criticized before
\citep{bran96a,saha97}. Indeed, if the Galactic absorption law applies
on average, the coefficient of the color term in equations (6) to (8)
would have to be $\sim$4, $\sim$3, and $\sim$2, respectively, which is
excluded at the 3-4 sigma level. The decisive proof against absorption
being the main source of the color variations is the fact that SNe Ia
become {\em brighter} in $I$ as their ($V$-$I$) colors become {\em
redder} \citep{saha99}, which is here confirmed.

\placefigure{fig5}

The color correction has no effect on $H_0$ derived below, because the
calibrators and the fiducial sample have closely the same colors
(cf. Table 4).

\subsection{Combining decline rate $\Delta m_{15}$ and color ($B$-$V$)}

The colors ($B$-$V$) of the SNe Ia of the fiducial sample do not
correlate with the decline rates $\Delta m_{15}$. These two parameters
being orthogonal should hence be combined for an optimum
homogenization of the blue SNe\,Ia. If the residuals $\delta
m$\,=\,$m_{obs}$-m$_{fit}$ of the fiducial sample are fit
simultaneously by a least-squares solution for two free parameters
linear in $\Delta m_{15}$ and ($B$-$V$) one obtains
\begin{eqnarray}
\!\!\!\!\!\!\!\delta m_B^{\rm corr}&\!\!\!=\!\!\!\!&0.44_{\pm0.13}\,(\Delta m_{15}-1.2)+2.46_{\pm0.46}\,[(B-V)+0.01]-28.400_{\pm0.161},\,\sigma_B=0.129\\
\!\!\!\!\!\!\!\delta m_V^{\rm corr}&\!\!\!=\!\!\!\!&0.47_{\pm0.11}\,(\Delta m_{15}-1.2)+1.39_{\pm0.40}\,[(B-V)+0.01]-28.391_{\pm0.143},\,\sigma_V=0.129\\
\!\!\!\!\!\!\!\delta m_I^{\rm corr}&\!\!\!=\!\!\!\!&0.40_{\pm0.13}\,(\Delta m_{15}-1.2)+1.21_{\pm0.43}\,[(B-V)+0.01]-28.105_{\pm0.171},\,\sigma_I=0.122
\end{eqnarray}
The coefficients are similar and their errors are equal to those
appearing in equations (3) to (8). They are also in statistical
agreement with those proposed by \citet{trip98} and \citet{trip99}. A
Fisher F-test for an additional term \citep{bevi92} shows that the
inclusion of a second parameter is significant at the 99.9\,percent
level. The scatter is now reduced to $\stackrel{_<}{_\sim}$0\magn13
which is significantly less than from either $\Delta m_{15}$ or
($B$-$V$) corrections alone. In fact the scatter is now of the same
order as an (optimistic) estimate of the combined observational errors
in $m$, $\Delta m_{15}$, and ($B$-$V$).

\subsection{Hubble Type}

A correlation between the SNe Ia luminosities and the {color of the
parent galaxy} has been pointed out by \citet{bran96b}. Instead of
galaxy color we consider here, following \citet{hamu95} and
\citet{saha97,saha99}, the Hubble type of the parent galaxy because
many of them have known Hubble types but no colors.

The left panel of Fig.\,6 shows the correlation of the residuals
$m_{obs}$-m$_{fit}$ on the Hubble type $T$ (from Table 1, column 2). A
weighted least-squares fit gives for the $V$ residuals, where the
effect is most pronounced,
$m_{obs}$-m$_{fit}$$\sim$(-0.017$\pm$0.013)\,$T$, which is only
moderately significant. Taken at face value it implies that SNe\,Ia in
E galaxies ($T$=-3) are on average fainter by $\Delta V$=0.14$\pm$0.10
mag than their counterparts in Sc spirals ($T$=5). The type dependence
is somewhat stronger for SNe\,Ia within 10\,000\,km\,s$^{-1}$ and
disappears at large distances; this can presently only be explained by
invoking a statistical fluke. The luminosity dependence on the stellar
population is presumably due to metallicity effects and/or different
structures of the progenitor white dwarfs.

After homogenization of the fiducial sample as to decline rate $\Delta
m_{15}$ and color ($B$-$V$) according to equations (9) to (11) any
significant dependence of the $m^{corr}$-m$_{fit}$ on $T$
disappears. The reason is that there is a clear correlation of the
decline rate $\Delta m_{15}$ and the Hubble type. However, correcting
the magnitude residuals by $T$ instead of $\Delta m_{15}$ is less
effective, leaving a larger scatter about the Hubble line.

\placefigure{fig6}
\placefigure{fig7}

\subsection{Galactocentric distance}

In view of the dependence of SNe Ia luminosities on the Hubble type it
is reasonable to ask whether the luminosities of SNe Ia depend on
their distance $r_{\rm offset}$ from the center of their host galaxy,
expressed in units ot the galaxy diameter $r_{25}$. The residuals
$m_{obs}$-m$_{fit}$ are plotted versus the relative galactocentric
distance $r_{\rm offset}/r_{25}$ in the left panel of Fig.\,7 for all
SNe\,Ia of the fiducial sample for which $r_{25}$ is known. The
interpretation of the formally significant fit of
$m_{obs}$-m$_{fit}$\,$\sim$\, (0.27$\pm$0.11)\,$r_{\rm
offset}$/$r_{25}$, in all three colors, requires some caution because
of the type mix and the uneven distribution in galactocentric
distance. For spirals alone the effect is insignificant. Also the
luminosity scatter does {\em not} significantly change with radial
distance. This is in variance with \citet{wang97}
and \citet{ries99} who suggested a larger luminosity scatter for
the inner SNe Ia than for the outer ones. Their results depend heavily
on the spectroscopically peculiar objects SN 1986~G, 1991~bg, and
1991~T, which happen to lie at small galactocentric distances; they
are excluded here (cf. Section 2). Once the magnitude residuals are
corrected for differences in decline rate and color through equations
(9) to (11), the remaining slope of 0.11$\pm$0.08 (cf. Fig.\,7, right
panel) becomes even more marginal. \citet{ries99}, discussing
absolute galactocentric distances, came to the same conclusion. In any
case the dependence of the luminosity on galactocentric distance has
zero effect on the value of $H_0$ derived below.
 
%------------------------------------------------------------------
\section{The value of the Hubble constant}

After correcting the magnitudes of the SNe Ia of the fiducial sample
for differences in the decline rate $\Delta m_{15}$ and the color
($B$-$V$) by means of equations (9) to (11), they define a Hubble
diagram with a much reduced scatter,
i.e. $\sigma_B$$\sim$$\sigma_V$$\sim$$\sigma_I$$\stackrel{_<}{_\sim}$0\magn13
(Fig.\,8). In fact the Hubble diagram is now so well defined that one
may ask if a stand on a specific model Universe is required for
relatively local SNe with
$v$\,$\stackrel{_<}{_\sim}$\,30\,000\,km\,s$^{-1}$. The situation is
visualized in a differential Hubble diagram (Fig.\,9). Three different
model Universes are fitted to the data: 

1) A flat universe with
$\Omega_M$\,=\,1.0 (q$_0$=0.5; Sandage 1961, 1962). Taking the
corresponding Hubble line and inserting the absolute magnitudes of the
calibrators, corrected from equations (9) to (11) to be $M_B^{\rm
corr}$=-19\magn48$\pm$0\magn07, $M_V^{\rm
corr}$=-19\magn47$\pm$0\magn06, and $M_I^{\rm
corr}$=-19\magn19$\pm$0\magn09,  leads to a $\chi^2$ solution with
$\chi_{\nu,B}^2$=0.696 ($\sigma_B$=0\magn130) and
$H_0$(B)=60.2$\pm$2.1. The values in $V$ and $I$ are very similar
(60.1/60.0). The data give a somewhat better fit (at the 1$\,\sigma$
level) on the assumption that $H_0$ is locally higher than the
asymptotic value, i.e. $H_0$(v$<$10\,000\,km\,s$^{-1}$)=60.8
($\sigma_B$=0.137) and $H_0$(v$>$10\,000\,km\,s$^{-1}$)=59.6
($\sigma_B$=0.118) \citep{saha99,tamm99}. The difficulty with this
solution is that the observational evidence speaks against a matter
density as high as $\Omega_M$\,=\,1.0 (for reviews cf., e.g., Bahcall
1997; Dekel et al. 1997; Tammann 1998). 

2) An open Universe with
$\Omega_M$\,=\,0.2 (q$_0$=0.1; Sandage 1961, 1962). As compared to
case 1) the $\chi_{\nu}^2$ solution gives a better fit with
$\chi_{\nu,B}^2$=0.633 ($\sigma_B$=0\magn124) and
$H_0$(B)=60.3$\pm$2.0. The solutions for $V$ and $I$ are again very
similar (60.2/60.1). The fit could still be improved with a smaller
$\Omega_M$, but $\Omega_M$=0.2 is about as low as independent
observations allow. The main difficulty here is, however, that the CMB
fluctuations strongly suggest a flat universe with $\Omega_{\rm
total}$=1.0 \citep{melc99,maci00}. 

3) A flat Universe with
$\Omega_M$=0.3, $\Omega_\Lambda$=0.7 (q$_0$=-0.55). This is the model
favored when including high-redshift SNe Ia out to $z$$\sim$0.8
\citep{perl98,perl98b,perl99,ries98b,schm98}. The fit to the fiducial sample by means
of equation (1) is with $\chi_{\nu,B}^2$=0.631,
$\chi_{\nu,V}^2$=0.802, $\chi_{\nu,I}^2$=0.589 ($\sigma_B$=0\magn124,
$\sigma_V$=0\magn123, $\sigma_I$=0\magn116) somewhat better than with
the two previous cases. However, an F-test shows that the improvement
of the total $\chi_{\nu}^2$ from three colors as compared to the case
with $\Omega_M$=1, $H_0$=constant, has a probability of not being a
result of chance of only $\sim$70\,percent. The corresponding values
of the Hubble constant are $H_0(B)$=61.0$\pm$2.1,
$H_0(V)$=60.9$\pm$1.8, and $H_0(I)$=60.7$\pm$2.6.

Not surprisingly, the specific choice of the cosmological model has
only a minor effect on $H_0$. In all three cases $H_0$ lies
within 60 $<$ $H_0$ $\le$ 61. Taking case 3) as the most consistent
solution, a value of
\begin{eqnarray}
   H_0&=&60.9\pm4.0\;\;\;(2\,\sigma\;\;{\rm error})
\end{eqnarray}
is found. The errors account only for the statistical error of the
absolute magnitude calibration (Table 3) and for the scatter about the
Hubble line. 

The solution in equation (12) is still affected by systematic
errors. Sources of systematic errors and our estimate of their sizes
are:

(1) The zeropoint of the DoPHOT photometry, on which six of the eight
Cepheid distances in Table 3 are based, was independently checked by
G00 as discussed in Section 2.2. They suggest that the DoPHOT
magnitudes are too faint by 0\magn04. On the other hand, a new
zeropoint determination of the DoPHOT photometry shows that the
previous zeropoint is still too bright by 0\magn02
\citep{saha00}. Correspondingly, a zeropoint error of $\pm$0\magn04 is
allowed for.

G00 have proposed a total reduction of the distances of the
Sandage/Saha team by 0\magn17 on average. Yet it was shown in Section
2.2 that if G00 had used only those Cepheids found in common by both
ALLFRAME and DoPHOT photometry (as appears to have been done for all
the other galaxies studied by the MFK team) they would have obtained
distances that are consistent within the statistical errors with those
in Table\,3. The smaller distances of G00 depend on the additional
Cepheids found by ALLFRAME alone.

(2) Photometric blending of the Cepheids may lead to a systematic
underestimate of the distances \citep{moch99}. \citet{stan99} have
proposed that the effect, increasing with distance, can amount to
0\magn3 at 20 Mpc. Counter-arguments by \citet{gibs00a}, also supported
by \citet{ferr00}, were judged to be weak \citep{pacz00}. In general,
it may be noted that the discovery mechanism for Cepheids favors
large-amplitude Cepheids, while the amplitudes of blended Cepheids are
reduced in function of the importance of the blend; moreover,  many
small-amplitude Cepheids have been excluded from our discussion
because of their less convincing light curves. Indeed, a detailed
analysis \citep{saha00} of the effect in $V$ and $I$ of skewed error
distributions from object confusion and surface brightness
fluctuations reveals that the distance modulus of NGC 4639 has been
underestimated by only 0\magn07. This galaxy is the most distant one
in Table 3, and the effect must be smaller for the others. An average
distance modulus increase of 0\magn03$\pm$0\magn03 seems reasonable.

(3) The adopted zeropoint of the Cepheid PL relation of
($m$-$M$)$_{\rm LMC}$=18.50 is likely to be too small by
0\magn06$\pm$0\magn10
\citep{fede98,grat98,mado98,feas99,walk99,gilm99}. Smaller LMC moduli
suggested on the basis of statistical parallaxes of RR Lyr stars and
red-giant clump stars depend entirely on the sample selection and on
the absence of metallicity and evolutionary effects, respectively. The
higher LMC modulus will increase all moduli by 0\magn06$\pm$0\magn10.

(4) The effect of metallicity variations on the Cepheid distances is a
long-standing problem. Much progress has been made on the theoretical
front. \citet{saio98} and \citet{bara98} have evolved Cepheids through
the different crossings of the instability strip and have investigated
the pulsational behavior at any point. The resulting, highly
metal-independent $M({\rm bol})$ have been transformed into PL
relations at different wavelengths by means of detailed atmospheric
models; the conclusion is that any metallicity dependence of the PL
relations is nearly negligible \citep{sand98,alib99}. \citet{bono98}
have suggested a much stronger metallicity dependence, but their
conclusions depend entirely on the precarious treatment of convection
at the red boundary of the instability strip.

From an observational point of view not even the sign of the
metallicity effect on the luminosity is unanimously accepted. Based on
[O/H] measurements the calibrating galaxies of Table\,3 have a range
of metallicities \citep{kenn99}. Allowing for this effect, G00 have
suggested that their distances should be increased by 0\magn07 on
average. \citet{kenn98} and \citet{feas99} recommend, on the other
hand, that for the present no metallicity correction should be
applied, but that for an uncertainty of $\stackrel{_>}{_\sim}
0.1\,$[Fe/H]$^{-1}$ \citep{feas99} should be allowed for. As a
compromise it is adopted here that the distance moduli in Table 3 are
underestimated by 0\magn04$\pm$0\magn10 on average.

(5) There is the general trend of an incomplete Cepheid sampling near
the photometry threshold to yield too short distances
\citep{sand88,lano99}. According to \citet{nara98} and \citet{mazu99}
this effect has caused a distance underestimate of $\sim$0\magn3 in the
extreme case of M100 \citep{ferr96}. This bias can be minimized by
introducing a period cutoff, if the data allow so, at an appropriate
low-period limit. Since the Cepheid moduli in Table 3 were derived in
cognizance of the selection bias, the net effect on the luminosity
calibration in Table 3 is most likely less than 0\magn1. Here a
systematic error of 0\magn05$\pm$0\magn05 is allowed for.

(6) An overestimate of the absorption of the
calibrating SNe\,Ia will lead to a spuriously low value of $H_0$, and
vice versa. The opposite holds for over-corrected SNe\,Ia in the
field. Yet a differential error of the extinction can be excluded at
the level of 0\magn01, because the calibrating SNe and those of the
fiducial sample have nearly identical corrected colors
(cf. Table\,4). An error of the adopted {\em mean} color of SNe\,Ia has no
effect on the distance scale because calibrating and field SNe\,Ia
would be equally affected by the resulting change of the
absorption. Concurrently the close agreement of $H_0$ from $B$, $V$,
and $I$ data speaks against hidden absorption problems. Absorption
corrections can therefore not affect $H_0$ by more than $\pm$ 3
percent.

A ratio of $R_V$=$A_V$/$E$($B$-$V$)=3.3 was adopted
here. With $R_V$=3.0 the calibrators would become fainter by 0\magn038
on average, but the fiducial sample only by 0\magn014. Therefore,
while prefering a value of $R_V$=3.3, a systematic uncertainty of the
distances of $\pm$0\magn02 should be allowed for.

It may be noticed that the nine SNe\,Ia of the fiducial sample with
Galactic absorptions $A_V$$>$0\magn2 have brighter absolute magnitudes
$M_B^{60}$ (Table\,1) than their 26 counterparts with smaller Galactic
absorption. The mean difference amounts to 0\magn21$\pm$0\magn04 and
is somewhat smaller in $V$ and $I$, as is to be expected if the
Galactic extinction corrections due to \citet{schl98} were
somewhat too large for large values. If the nine strongly corrected
SNe\,Ia are excluded, the Hubble line in Fig.\,3 shifts faintwards by
0\magn05$\pm$0\magn04, and $H_0$ is reduced by 2 percent. Furthermore
the scatter in Fig.\,8 shrinks to $\sigma_B$=0.114.

Finally, if the seven somewhat red SNe\,Ia in Table\,5, which were
excluded here on the assumption of internal absorption, were
intrinsically red and hence included in the present analysis, they
would slightly affect equations (9) to (11) and decrease $H_0$ by one
percent. If they had been included after corrections for absorption,
they would increase $H_0$ by much less than one percent.

(7) The coefficients of the $\Delta m_{15}$-term in equations (9) to
(11) carry a random error of $\pm$0.13. This is to be multiplied with
the mean difference $<$$\Delta m_{15}$$>$(fiducial
sample)$-$$<$$\Delta m_{15}$$>$(calibrators)=0.17 to give a systematic
error of the distances of $\pm$0\magn02. 

\citet{phil99} have pointed out that the decline rate $\Delta
m_{15}$ is slightly affected by absorption. If this effect had been
applied, the differential decline rate between calibrators and field
SNe\,Ia would be changed by $\delta \Delta m_{15}$\,=\,0.008 with
vanishingly small effect on $H_0$.

(8) The redshift velocities were corrected for the CMB dipole motion on
the assumption that the co-moving volume extends to
3000\,km\,s$^{-1}$. If instead the volume size was varied between 2000
and 10\,000\,km\,s$^{-1}$, it would affect $H_0$ by less than one
percent, as stated in Section 2.1. G00 have considered a very specific
local flow model and concluded that in this case $H_0$ would be
increased by two percent. This is taken as indication that the
deviations from pure Hubble flow could influence $H_0$ by hardly more
than $\pm$\,2\,percent.

Most of the systematic errors, which are assumed to give 90-percent
margins, tend to increase the true distance scale. If they are added
linearly, $H_0$ in equation (12) would be reduced by a factor
(0.91$\pm$0.07); if they are added in quadrature, instead, the
reduction factor becomes (0.96$\pm$0.08). Multiplying the latter with
equation (12) gives
\begin{eqnarray}
   H_0&=&58.5 \pm 6.3\;\;\;(2\,\sigma\;\;{\rm error}),
\end{eqnarray}
which is finally taken as our most probable value of the Hubble
constant as inferred from Cepheid-calibrated SNe\,Ia, including random
and systematic errors.

\section{Alternative solutions}

The second-parameter problem in finding $H_0$ could be avoided
altogether, if the distant SNe Ia and the local calibrators had
(nearly) identical second parameters. This can be approximated by
choosing a suitable subset of the fiducial sample. Excluding the 14
SNe Ia of the fiducial sample with the largest $\Delta m_{15}$ values
($\Delta m_{15}$$\ge$1.3), one is left with a rest sample of 21
objects with $<$$\Delta m_{15}$$>$=1.08$\pm$0.02, i.e. the same
as for the calibrators.

Fitting the 21 (uncorrected) SNe Ia to equation (1) by a 
$\chi^2$ solution and inserting the weighted absolute magnitudes of the
calibrators in Table 3 gives $H_0(B)$=58.8,
$H_0(V)$=58.8, and $H_0(I)$=59.9. This solution is
not yet fully satisfactory, because on average the calibrators are now
redder by $\Delta (B_0-V_0) = 0.030 \pm 0.018$ and the
Hubble type of their parent galaxies is still later by $\Delta T$\,=\,2.9$\pm$
0.7. But it should be noted that the result is rather close to the
corresponding, fully corrected value of $H_0$=60.9 in equation (12).\\
As the sample of known blue SNe Ia with
\( v\stackrel{_<}{_\sim}30\,000\) km\,s$^{-1}$ will increase, this
alternative solution will become more rigorously applicable and
circumvent all corrections for second parameters.

Another alternative solution is given by confining the analysis to SNe
Ia in spirals. Seven of the eight adopted calibrators lie in
spirals. The population assignment of SN 1972\,E is ambiguous because
its parent galaxy is of Hubble type Am;  the inclusion here of SN
1972\,E does not affect the result. Nineteen of the
SNe Ia of the fiducial sample after 1985 are in
spirals ($T$$\ge$0). Combining them with the calibrators and fitting them to
equation (1) yields $H_0(B)$=59.0, $H_0(V)$=59.1, and
$H_0(I)$=59.1. The spirals-only solution is remarkable
because the mean solution of $H_0$=59.1 is smaller by only
3\,\% than the solution fully corrected for $\Delta m_{15}$ and
(B$_0$-V$_0$) in equation (12).

Thus both the exclusion of fastest-declining SNe and the restriction
to only SNe in spirals provide useful simple approximations
for the determination of $H_0$.

\section{Conclusions}

A fiducial sample of 35 well observed, blue (Branch-normal) SNe\,Ia
with minimum absorption in their parent galaxies and with
1200\,$<$\,$v$\,$\stackrel{_<}{_\sim}$\,30\,000\,km\,s$^{-1}$ has been
compiled from the literature. For 29 SNe\,Ia also I-magnitudes are
known. The intrinsic color of SNe\,Ia is found to be
($B$-$V$)=-0\magn012$\pm$0.008, ($V$-$I$)=-0.276$\pm$0.016 from objects in
E/S0 galaxies and outlyers in spirals; the scatter in color of
$\sigma$=0\magn05 for individual objects is surprisingly small.

The fit of the SNe\,Ia sample, although quite local
($z$$\stackrel{_<}{_\sim}$0.1), to a Hubble line is somewhat dependent
on the cosmological model adopted. Among various models, the
marginally best fit is obtained for a flat Universe with
$\Omega_M$=0.3, $\Omega_\Lambda$=0.7, which is presently also favored
by external data. In this case the scatter about the Hubble line is
$\sigma$=0\magn21-0\magn16, depending on the pass band. But the
magnitude residuals correlate with decline rate $\Delta m_{15}$ and
color ($B$-$V$). If the dependency on either of these parameters is
removed, the scatter is reduced to $\sigma$=0\magn19-0\magn14. A
simultaneous correction for $\Delta m_{15}$ and color ($B$-$V$)
decreases the scatter to $\sigma$$\stackrel{_<}{_\sim}$0\magn13 in all
three colors. This illustrates the unique potential of SNe\,Ia as
distance indicators.

If the SNe\,Ia of the fiducial sample are assigned the absolute
magnitudes $M_B$, $M_V$, and $M_I$ of eight local SNe\,Ia whose
Cepheid distances are known, one obtains closely the same value of
$H_0$=58.3$\pm$2.0 for the three passbands. This solution is remarkably
robust against various assumptions and corrections. Bootstrapping the
calibrators by excluding two or three objects, does not change their
mean absolute magnitude. The eight calibrators lie essentially all in
spirals and they have correspondingly a somewhat smaller average
$\Delta m_{15}$ than the fiducial sample. If they are adjusted to the
latter's mean $\Delta m_{15}$, $H_0$ is increased by 4 percent to
$H_0$=60.9$\pm$2.0. A correction for ($B$-$V$) does not change this
value because the calibrators and the fiducial sample have closely the
same mean color. Any remaining dependencies on Hubble type and
galactocentric distance are insignificant and have no effect on
$H_0$. Alternative solutions considering either SNe\,Ia only in
spirals or SNe\,Ia with small $\Delta m_{15}$ --- making the sample
more similar to the calibrators --- yields $H_0$$\approx$59. The
effect of systematic errors is mainly one-sided (photometric
zeropoint, skewness of photometric errors, the LMC zeropoint of the PL
relation, selection effects of Cepheids, and possibly metallicity
effects), leading to too high values of $H_0$. If they are taken into
account one obtains at the 90 percent confidence level
$H_0$=58.5$\pm$6.3.

External results on $H_0$ from SNe\,Ia should be compared with
$H_0$=60.9$\pm$2.0 from equation (12), because the subsequent
corrections for systematic errors have not been applied yet in other
papers. The results of \citet{saha99} and \citet{trip98} are
indistinguishable from the above result despite of the re-definition
of the ``fiducial sample'' here, of variations of other input data,
and of independent corrections for $\Delta m_{15}$ and  ($B$-$V$). The
value of $H_0$=62($\pm$2) derived by \citet{trip99} from a sample of
six calibrators and 13 field SNe\,Ia in spirals and fully corrected
for $\Delta m_{15}$ and ($B$-$V$)  is in satisfactory statistical
agreement with equation (12), too.

The somewhat higher value of $H_0$=63.9$\pm$2.2 by \citet{sunt99},
based on only five of the calibrators in Table 3, is fully explained
by their adopted steeper luminosity dependence on $\Delta m_{15}$,
which is closely the same as was derived by \citet{phil99}, who
included spectroscopically peculiar SNe\,Ia like SN\,1986\,G, 1991\,T,
1991\,bg, and 1995\,ac. Had Suntzeff's et al. (1999) $\Delta
m_{15}$-correction been applied to our fiducial sample, we would have
obtained $H_0$=59.7 for the SNe\,Ia with $\Delta m_{15}$$<$1.2
($n$=17), and $H_0$=64.6 for the ones with $\Delta m_{15}$$\ge$1.2
($n$=18). Since seven of the eight calibrators fall into the first
category the lower value of $H_0$ is more nearly correct. The
3-$\sigma$ discrepancy demonstrates that their $\Delta
m_{15}$-correction introduces an over-correction, making the
calibrators with a small mean $\Delta m_{15}$ too faint and the field
SNe\,Ia with large $\Delta m_{15}$ too bright. This is the price to be
paid if one attempts to derive a global correction formula which
encompasses also peculiar objects that are extraneous of the SNe\,Ia
sample to which it should be applied. --- G00 have followed closely
the line of \citet{sunt99}, but they have reduced in addition the mean
absolute magnitude of the calibrators by $\sim$0\magn2 on the basis of
quite objectionable additional Cepheids (cf. Section 2.2); their high
value of $H_0$=68 (based on six calibrators) then follows by
necessity. The repetition of one or two calibrating galaxies with the
forthcoming Advanced Camera System on HST will bring a definitive
decision on the merits of the additional Cepheids of G00, which were
rejected here because their photometry relies so far only on ALLFRAME.

\citet{jha99} have derived $H_0$=64.4$\pm$5.4 from a sample of 42
distant SNe\,Ia and only four Cepheid-calibrated SNe\,Ia. Seven of
these objects have not been used here as either being too red or
having peculiar spectra (SN 1992\,K, 1995\,ac). The SNe\,Ia are
standardized by a ``multi-color light curve shape'' (MCLS) method
giving simultaneously the absorption and a light-curve parameter
substituting $\Delta m_{15}$ \citep{ries96,ries98b}. Unfortunately the
individual data are not given, and the discussion is restricted to $V$
magnitudes denying a control on the consistency of colors. In spite of
this, there is some question as to the grip of the MLCS light-curve
parameter. The calibrators as well as the distant SNe\,Ia have still a
magnitude scatter of $\sigma_V$=0\magn16 after the correction, which
is larger than $\sigma$=0\magn14 obtained by \citet{phil99} and
significantly larger than $\sigma_V$$\le$0\magn13 found here by means
of the corrections due to equation (10).

This discussion confirms previous investigations suggesting that if
one attempts to derive $H_0$ from SNe\,Ia to within
$\stackrel{_<}{_\sim}$10\,percent, a major stumbling block are the
second-parameter corrections. If the relatively mild $\Delta
m_{15}$-corrections found here can be agreed upon it is clear that
$H_0$ will settle close to 60 with a tendency for a small downward
correction due to various (small) systematic errors.

Future Branch-normal SNe\,Ia in the field, preferably with
$v$\,$\stackrel{_>}{_\sim}$\,3000\,km\,s$^{-1}$, will help to settle
the problem of light-curve shape corrections. The importance of these
corrections will decrease as it will become possible to define
sufficiently large sub-samples of field SNe\,Ia which have the same
mean second parameters as the calibrating SNe\,Ia. Additional field
SNe\,Ia are forthcoming \citep{nuge99}. Finally, the
advent of the new Advanced Camera System on HST will allow to
check the present Cepheid distances and to increase the number of
calibrating SNe\,Ia.

At present the evidence from SNe\,Ia is best reflected at the
90-percent confidence level by a value of the Hubble constant,
including random and systematic errors, of $H_0$=58.5$\pm$6.3.

\acknowledgments

We are grateful to B.\,Leibundgut for sharing some template-fitting
software, M.\,Hamuy for providing us with a set of I band
K-corrections, and B.\,Reindl for useful discussions. G.A.T. has
profited from a Workshop on Supernovae (June 1999) at the Aspen Center
for Physics. B.R.P. and G.A.T. acknowledge financial support of the
Swiss National Science Foundation.

%=================== REFERENCES ===============================

%=========================== FIGURE CAPTIONS =============================
\newpage

\figcaption[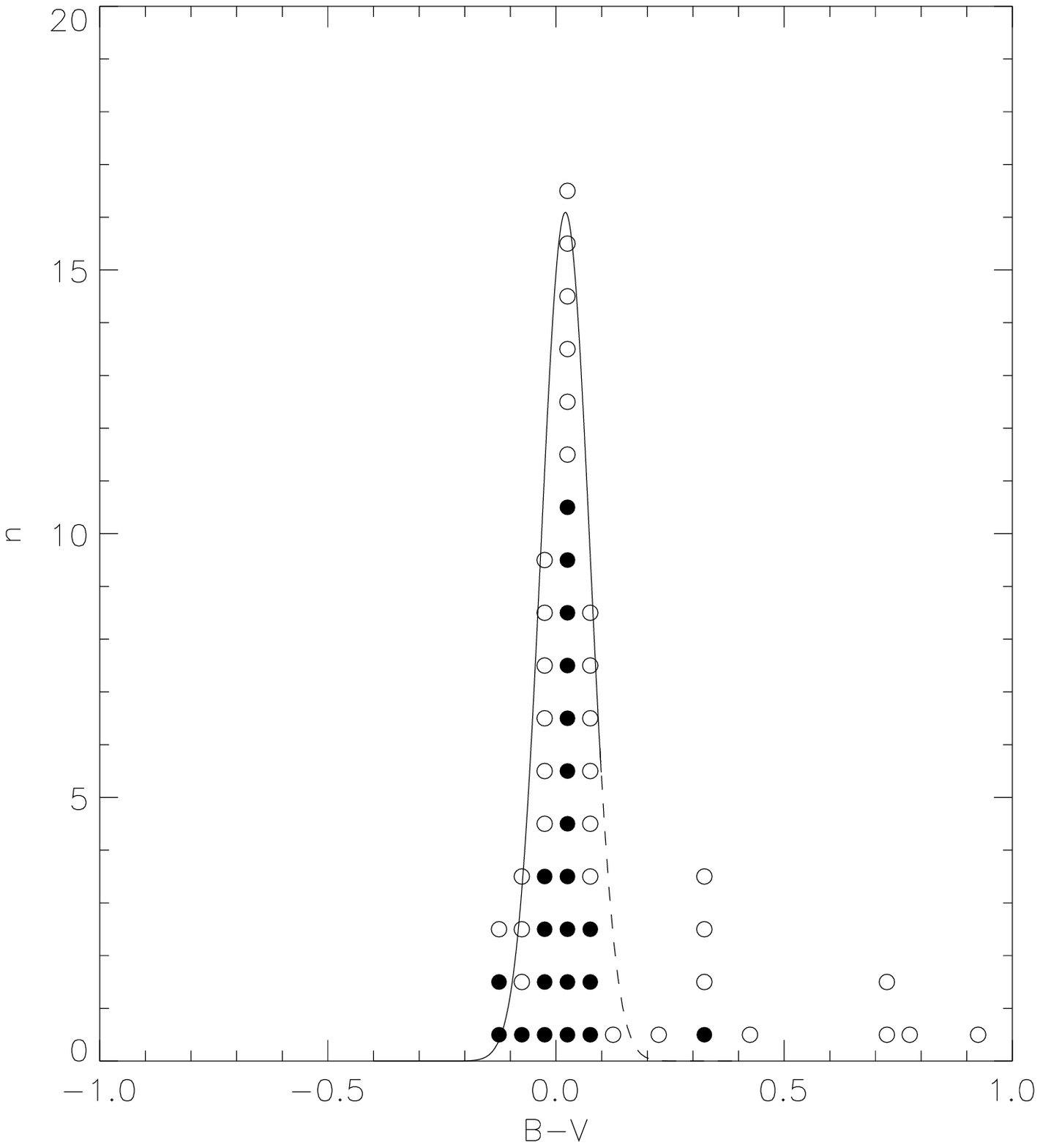]{The color distribution of all known SNe\,Ia
after 1985 with v$\stackrel{_<}{_\sim}$30\,000\,km\,s$^{-1}$. Open
symbols are for SNe\,Ia with v$<$10\,000\,km\,s$^{-1}$, closed symbols
are for more distant SNe\,Ia. The binned intervals embrace
$\Delta$(B-V)=0\magn05. A Gaussian fit to all SNe\,Ia with
(B-V)$\le$0\magn10 gives $<$B-V$>$=0.020, $\sigma_{B-V}$=0.053.
\label{fig1}}

\figcaption[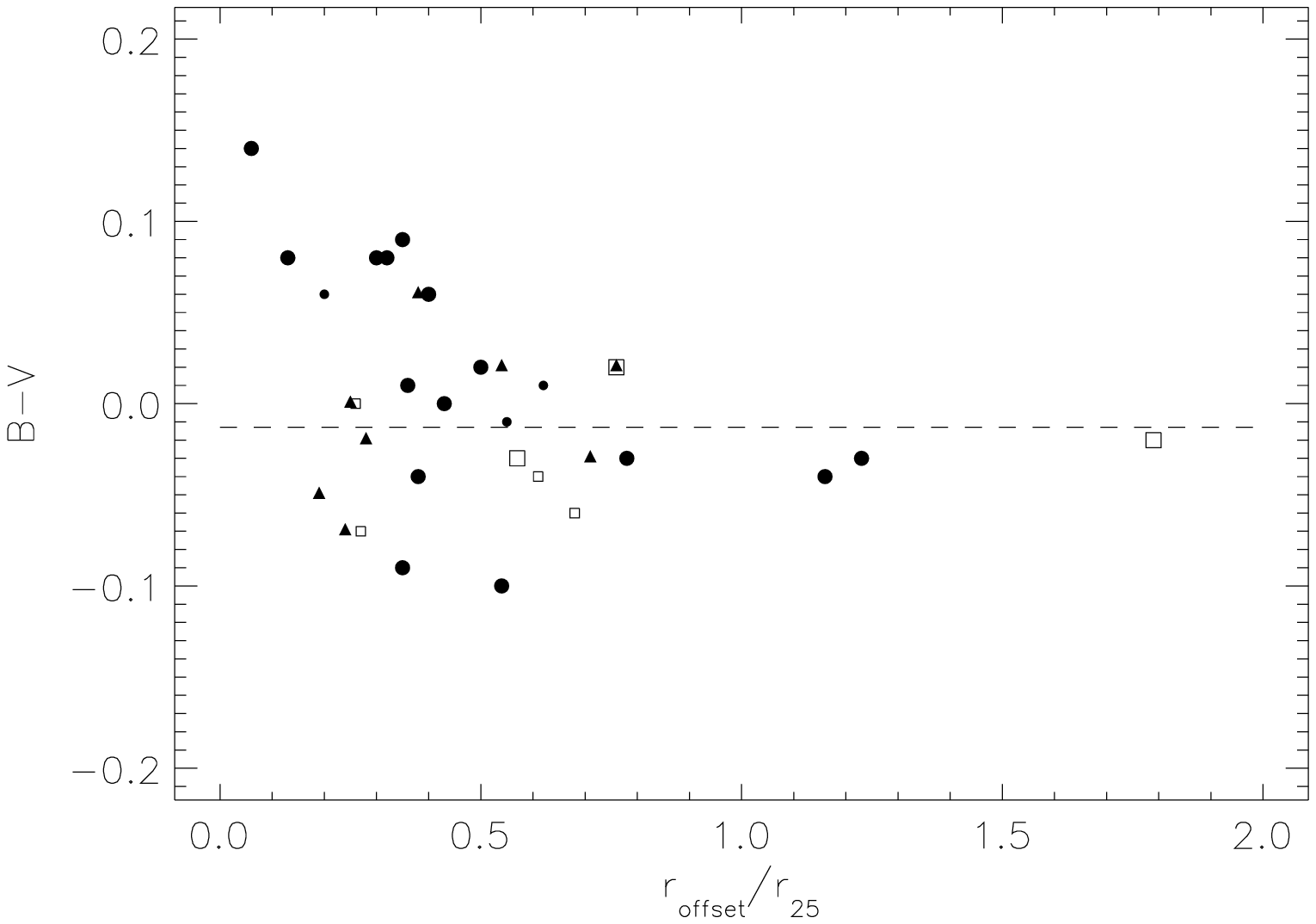]{Blue SNe Ia after 1985 plotted against the
relative radial distance r/r$_{25}$. Circles stand for spiral, squares
for E/S0 host galaxies. Small symbols represent SNe\,Ia with
observations starting eight days after $B$ maximum or later. Triangles
are the calibrators from Table\,3.
\label{fig2}}

\figcaption[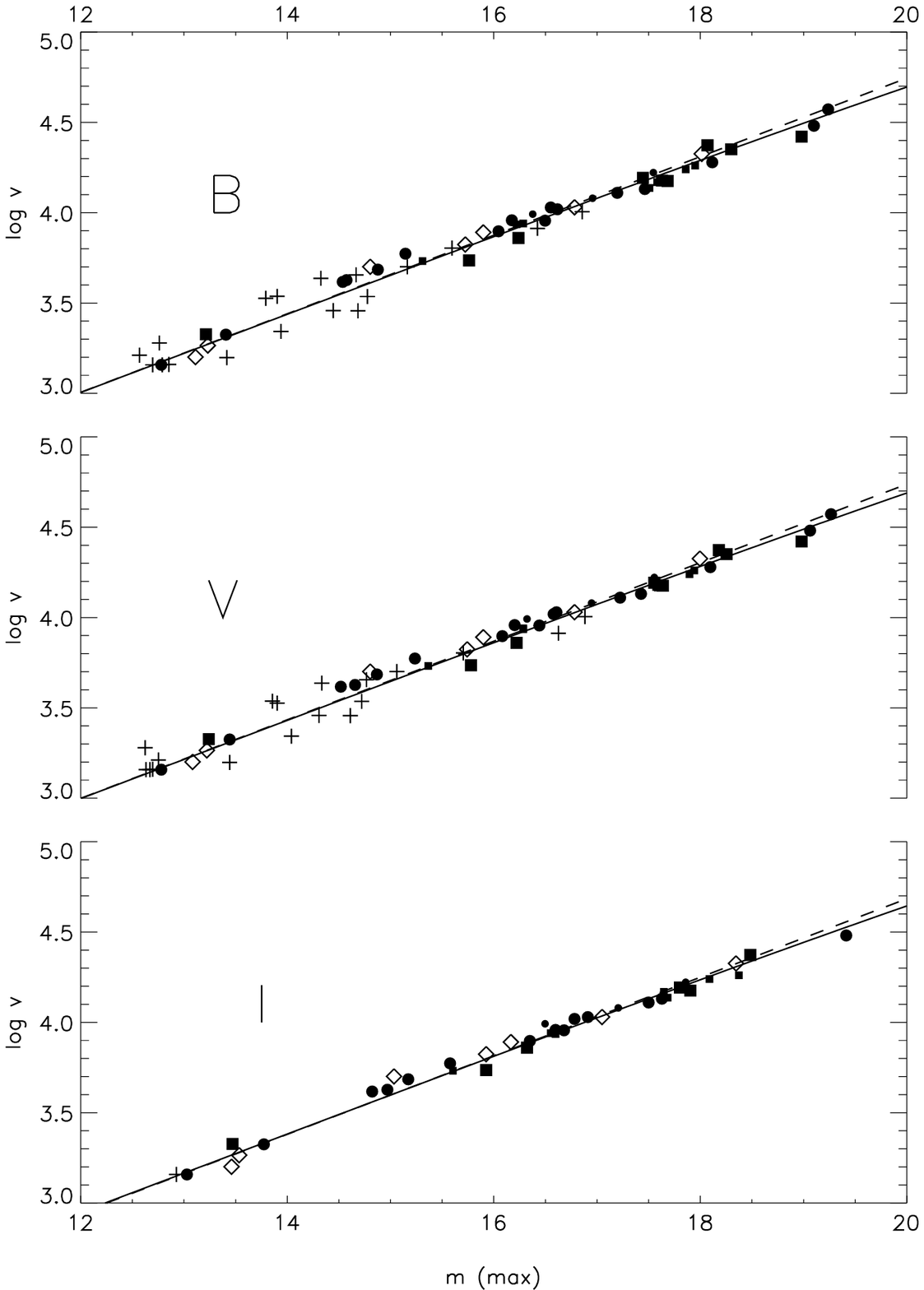]{The Hubble diagrams in $B$, $V$, and $I$ for the
fiducial sample. Circles are SNe\,Ia in spirals, squares in E/S0
galaxies. Small symbols are SNe\,Ia whose observations begin eight
days after B maximum or later. Solid lines are fits to the data
assuming a flat universe with $\Omega_M$=0.3 and $\Omega_\Lambda$=0.7;
dashed lines are linear fits with a forced slope of 0.2 (corresponding
approximately to $\Omega_M$=1.0 and $\Omega_\Lambda$=0.0.). Not
considered for the fits are the diamonds and the crosses representing
SNe\,Ia dereddened according to Table 5 or observed before 1985,
respectively.
\label{fig3}}
 
\figcaption[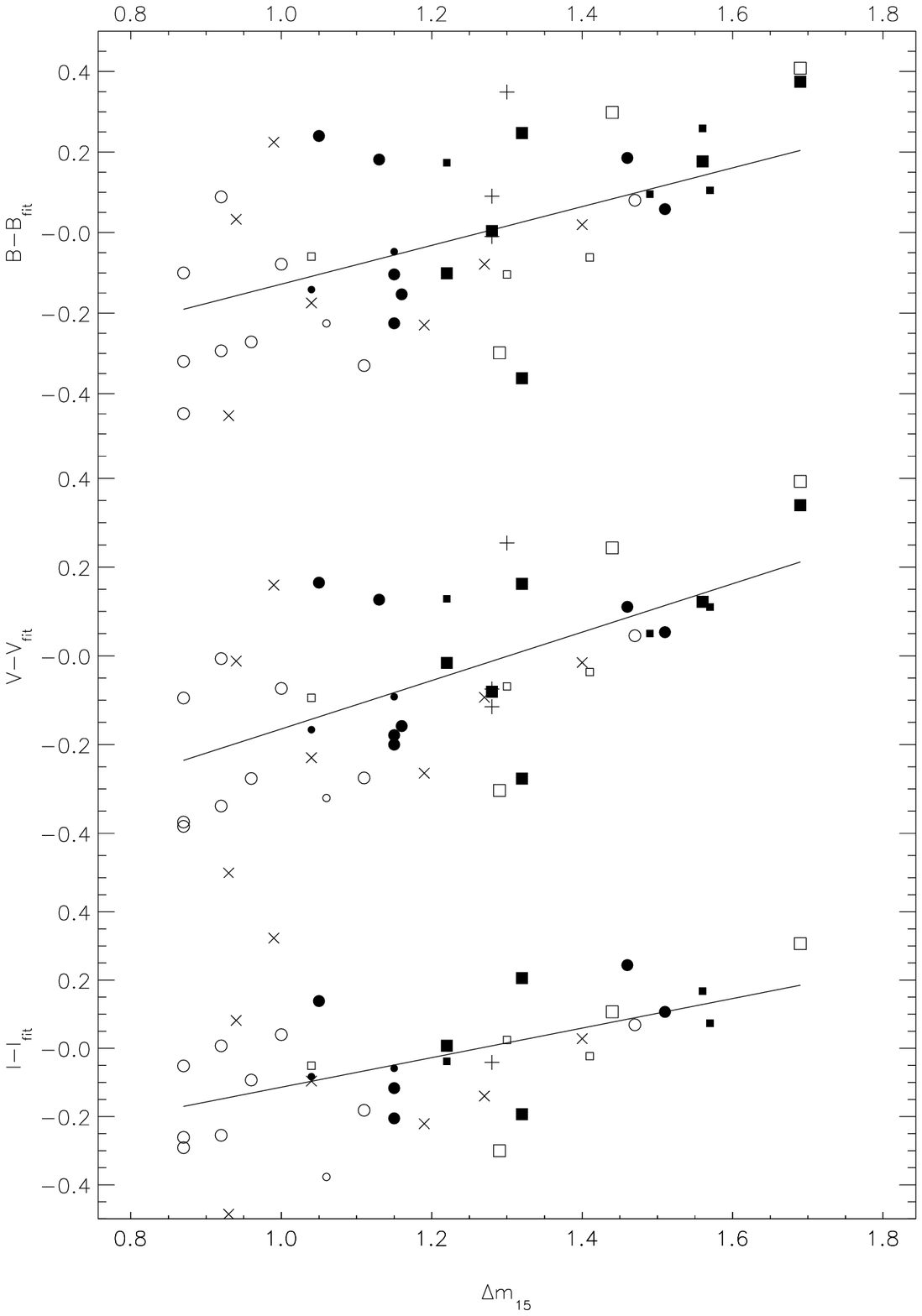]{Relative absolute magnitudes (i.e. residuals
from the Hubble line in Fig.\,3) for the SNe Ia of the fiducial sample
in function of the decline rate $\Delta m_{15}$. Circles are SNe\,Ia
in spirals, squares in E/S0 galaxies. Open symbols are SNe\,Ia with
1200 $<$ v $<$ 10\,000\,km\,s$^{-1}$, closed symbols are for the more
distant SNe\,Ia. Small symbols are SNe Ia whose observations begin
eight days after B maximum or later. Neither the SNe\,Ia before 1985
with known $\Delta m_{15}$ (shown as crosses) nor the seven blue, but
reddened SNe\,Ia (shown as $X$'s) are considered for the weighted
least-squares solutions (solid lines).
\label{fig4}}

\figcaption[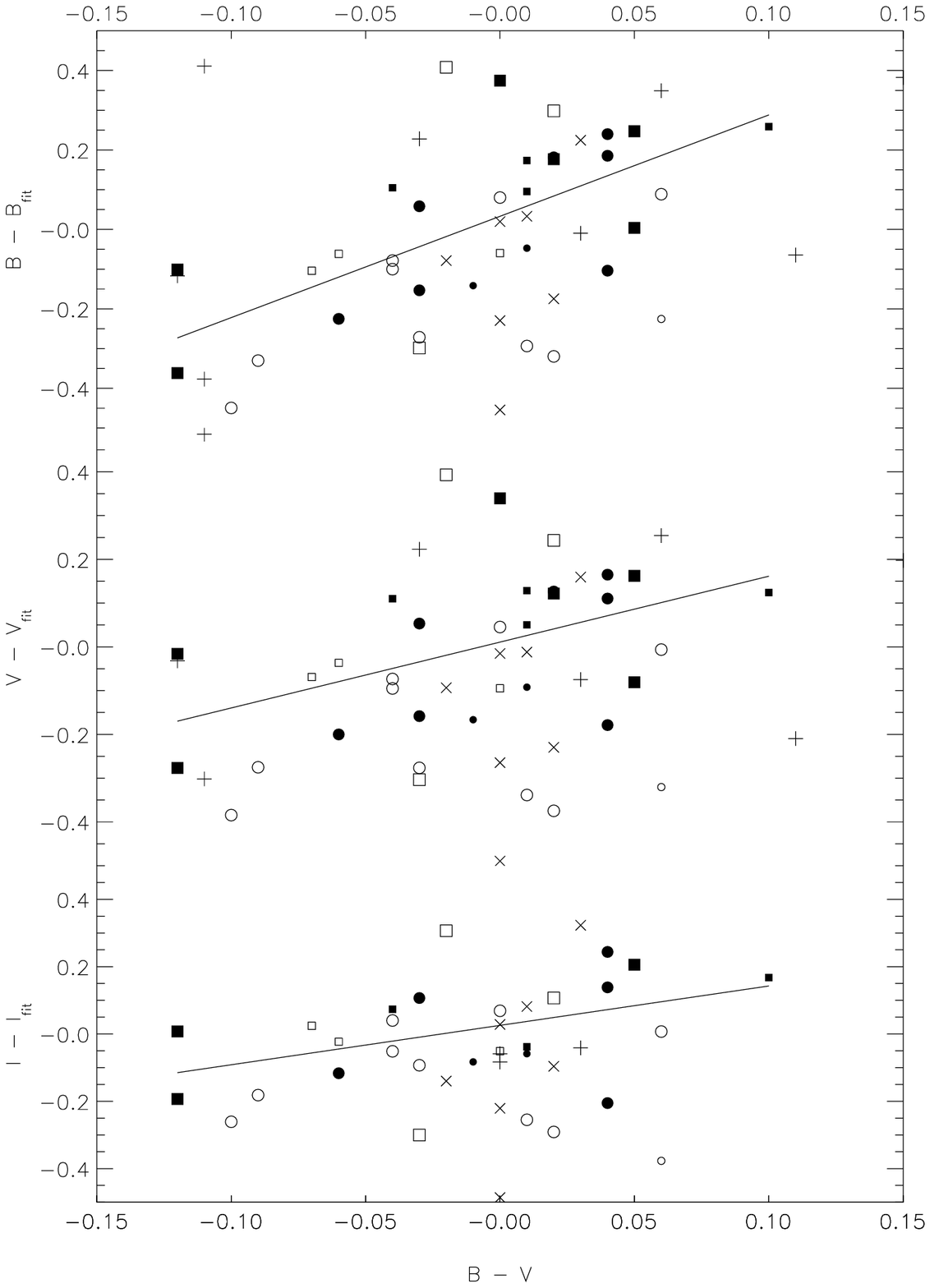]{Relative absolute magnitudes (i.e. residuals
from the Hubble line in Fig.\,3) for the SNe Ia of the fiducial sample
in function of their color ($B$-$V$). Symbols as in Fig.\,4. Neither
the few SNe Ia before 1985 (shown as crosses) nor the seven blue, but
reddened SNe\,Ia (shown as $X$'s) are considered for the weighted
least-squares solutions (solid lines).
\label{fig5}}

\figcaption[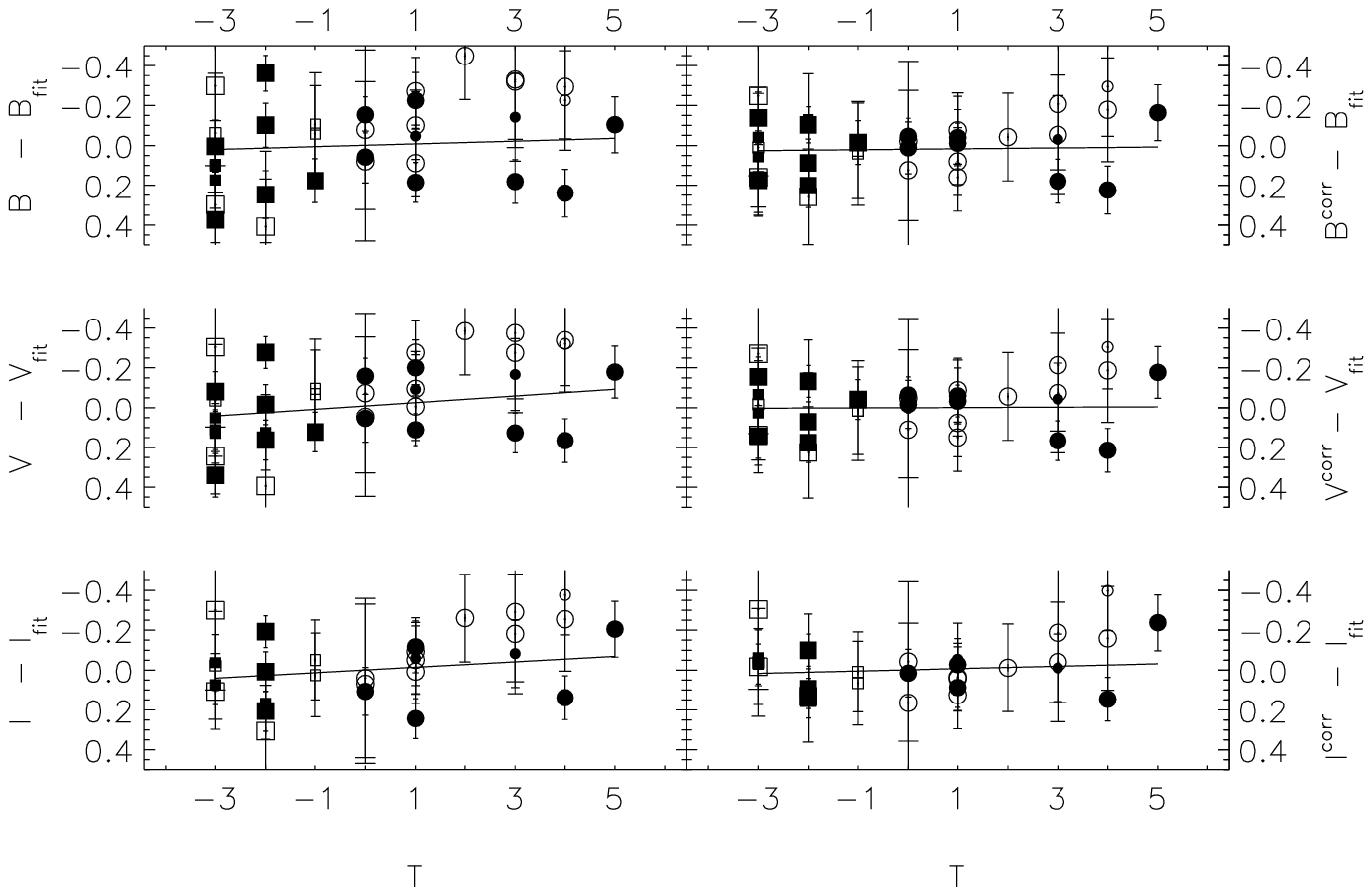]{Relative absolute magnitudes (i.e. residuals
from the Hubble line in Fig.\,3) for the SNe Ia of the fiducial sample
in function of their Hubble type $T$. Symbols as in Fig.\,4. Left
panel: $m_{obs}$-$m_{fit}$. Right panel: $m^{corr}$-$m_{fit}$,
i.e. after magnitude corrections according to equations (9) to (11).
\label{fig6}}

\figcaption[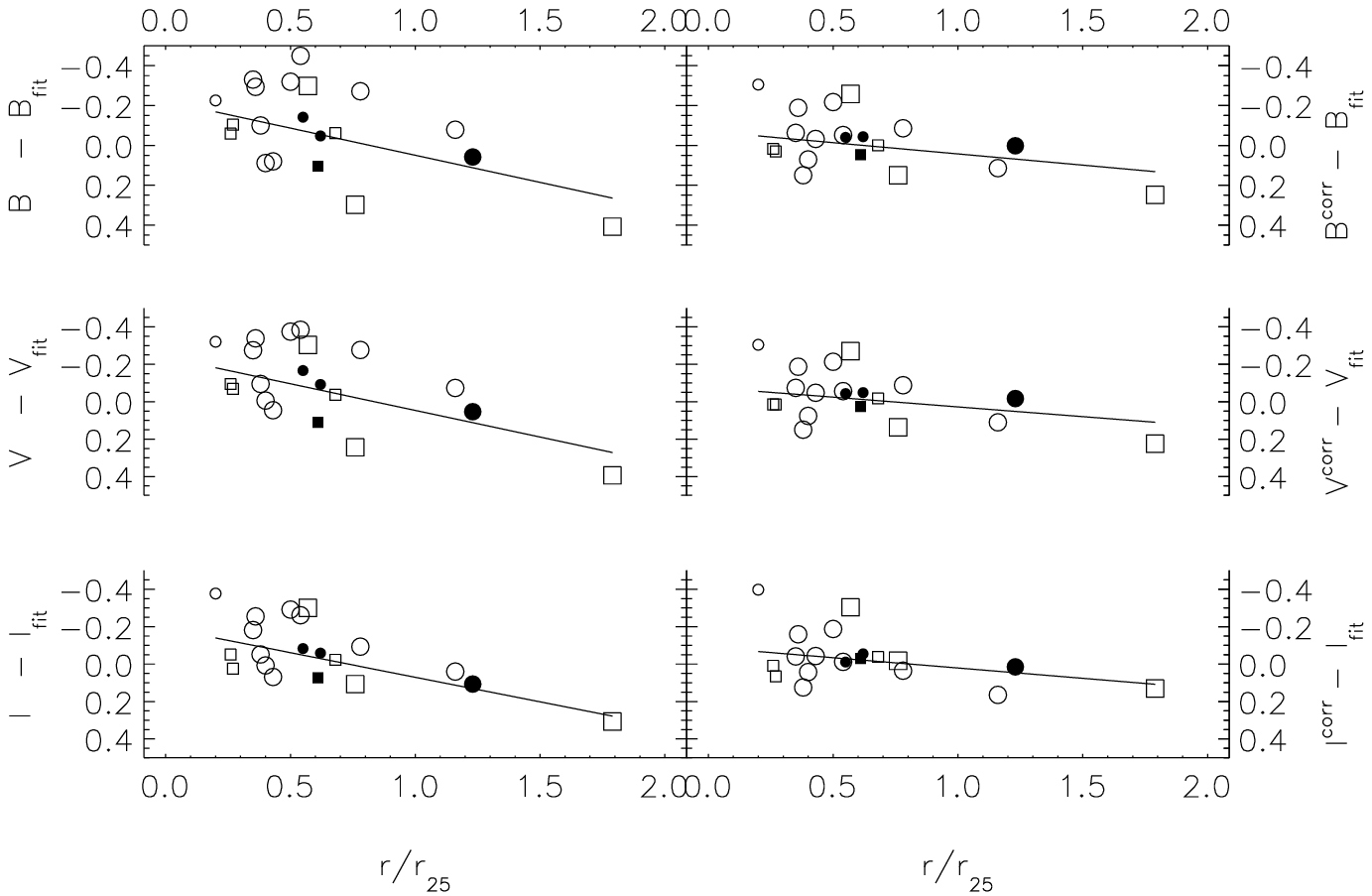]{Relative absolute magnitudes (i.e. residuals
from the Hubble line in Fig.\,3) for the SNe Ia of the fiducial sample
in function of their projected galactocentric distances
$r$/$r_{25}$. Symbols as in Fig.\,4. Left panel:
$m_{obs}$-$m_{fit}$. Right panel: $m^{corr}$-$m_{fit}$,
i.e. after magnitude corrections according to equations (9) to (11).
\label{fig7}}

\figcaption[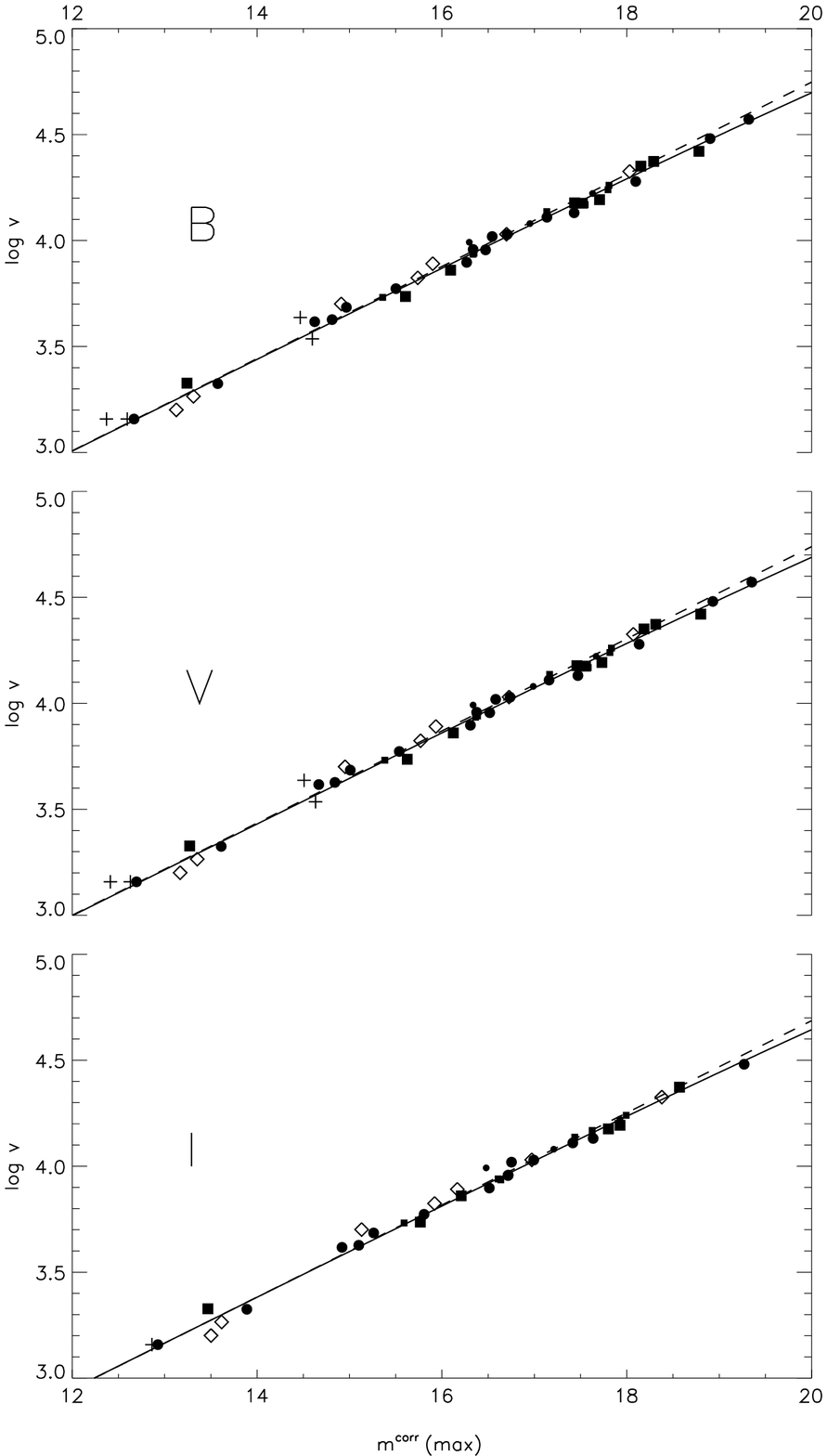]{The Hubble diagrams in $B$, $V$, and $I$ for the
35 (29) SNe\,Ia of the fiducial sample with magnitudes $m^{corr}$
(i.e. corrected according to equations (11) to (13)). Symbols as in
Fig.\,3. The solid line is for a model with $\Omega_M$=0.3,
$\Omega_\Lambda$=0.7, the dashed line for $\Omega_M$=1.0,
$\Omega_\Lambda$=0.0.
\label{fig8}}

\figcaption[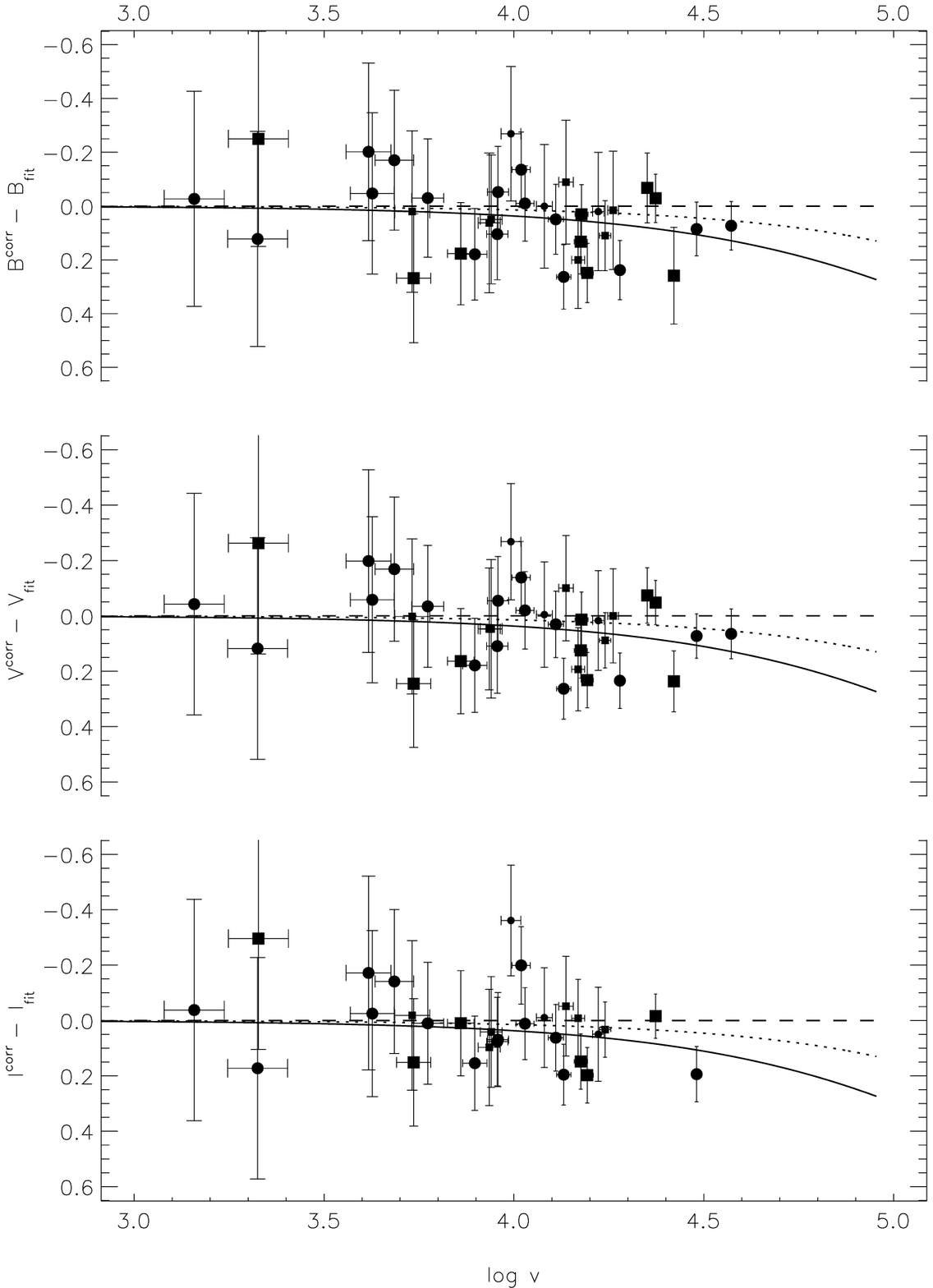]{Differential Hubble diagrams
($m^{corr}$-$m_{fit}$) vs $\log v$ in $B$, $V$, and $I$ for the 35
(29) SNe of the fiducial sample. Symbols as in Fig.\,3. The dashed
line is for a flat cosmological model with $\Omega_M$=1.0,
$\Omega_\Lambda$=0.0; the theoretical apparent magnitudes $m_{fit}$
correspond to this model. The full line is for a flat model with
$\Omega_M$=0.3, $\Omega_\Lambda$=0.7; the dotted line is for an open
universe with $\Omega_M$=0.2, $\Omega_\Lambda$=0.0.
\label{fig9}}

\newpage
\plotone{fig1.eps}
\begin{center} \small{fig.1} \end{center}

\newpage
\plotone{fig2.eps}
\begin{center} \small{fig.2} \end{center}

\newpage
\epsscale{0.9}
\plotone{fig3.eps}
\begin{center} \small{fig.3} \end{center}

\newpage
\plotone{fig4.eps}
\begin{center} \small{fig.4} \end{center}

\newpage
\plotone{fig5.eps}
\begin{center} \small{fig.5} \end{center}

\newpage
\plotone{fig6.eps}
\begin{center} \small{fig.6} \end{center}

\newpage
\plotone{fig7.eps}
\begin{center} \small{fig.7} \end{center}

\newpage
\epsscale{0.7}
\plotone{fig8.eps}
\begin{center} \small{fig.8} \end{center}

\newpage
\plotone{fig9.eps}
\begin{center} \small{fig.9} \end{center}

%==================== TABLES =============================================
\clearpage

\oddsidemargin=-1.0cm
\begin{deluxetable}{lrccccccrcccc}
\tablenum{1}
\tabletypesize{\tiny}
\tablewidth{0pt}
\tablecaption{\sc Photometric Parameters of Blue SNe Ia {(\,(B-V) $\le$ 0.20\,)} \label{tbl1}}
\tablehead{
\colhead{SN}        & \colhead{T}                 & \colhead{$\log v$}   &
\colhead{B}         & \colhead{V}                 & \colhead{I}          &
\colhead{A$_{V}$}   & \colhead{$\Delta $m$_{15}$} & \colhead{1$^{\rm st}$\,obs} &
\colhead{Ref.}      & \colhead{M$_B^{60}$}        & \colhead{M$_V^{60}$} & \colhead{M$_I^{60}$} \\
\colhead{(1)}&(2)&(3)&(4)&(5)&(6)&(7)&(8)&(9)$\;\;\;\;$&(10)&(11)&(12)&(13)}
\startdata
1937 C                  &  5 & 2.464(79) &  8.74(09) &  8.77(11) & \dots     & 0.05 & 0.87(10) & \dots & 4a&  ---*      &  ---*      & \dots      \\
1956 A                  &  3 & 3.160(79) & 12.58(30) & 12.41(30) & \dots     & 0.10 & \dots    &  +8:  & 2 & -19.33(50) & -19.50(50) & \dots      \\ 
1959 C                  &  5 & 3.526(71) & 13.60(20) & 13.72(20) & \dots     & 0.09 & \dots    &  +2:  & 2 & -20.15(41) & -20.03(50) & \dots      \\ 
1960 F                  &  5 & 3.072(41) & 11.49(10) & 11.43(15) & \dots     & 0.08 & 1.06(12) & \dots & 1 &  ---*      &  ---*      & \dots      \\  
1962 A                  & -2 & 3.804(39) & 15.56(40) & 15.68(40) & \dots     & 0.04 & \dots    & -15:  & 2 & -19.58(45) & -19.46(45) & \dots      \\  
1965 I                  & -1 & 3.211(79) & 12.27(30) & 12.47(30) & \dots     & 0.11 & \dots    & -10:  & 3 & -19.90(50) & -19.70(50) & \dots      \\ 
1966 K                  & -1 & 4.005(25) & 16.93(40) & 16.96(40) & \dots     & 0.06 & \dots    & -11:  & 2 & -19.22(42) & -19.19(42) & \dots      \\ 
1967 C                  &  5 & 3.198(79) & 13.19(30) & 13.22(30) & \dots     & 0.09 & \dots    &   0:  & 2 & -18.91(50) & -18.88(50) & \dots      \\  
1969 C                  &  5 & 3.537(70) & 13.72(30) & 13.67(30) & \dots     & 0.06 & \dots    &  +5:  & 3 & -20.08(46) & -20.13(46) & \dots      \\  
1970 J                  & -3 & 3.536(70) & 14.67(30) & 14.61(30) & \dots     & 0.26 & 1.30(..) & -12:  & 3 & -19.13(46) & -19.19(46) & \dots      \\  
1971 G                  &  1 & 3.343(79) & 13.76(25) & 13.87(25) & \dots     & 0.12 & \dots    & -17:  & 3 & -19.07(47) & -18.96(47) & \dots      \\  
1971 L                  &  3 & 3.279(79) & 12.48(25) & 12.33(25) & \dots     & 0.41 & \dots    &  -5:  & 3 & -20.03(47) & -20.18(47) & \dots      \\  
1972 E                  & 5: & 2.464(79) &  8.25(14) &  8.30(15) &  8.68(00) & 0.19 & 0.87(10) & \dots & 4a&  ---*      &  ---*      &  ---*      \\ 
1972 H                  &  3 & 3.458(79) & 14.31(40) & 14.16(40) & \dots     & 0.08 & \dots    &  +8:  & 3 & -19.09(56) & -19.24(56) & \dots      \\  
1972 J                  & -1 & 3.457(79) & 14.57(30) & 14.49(30) & \dots     & 0.15 & \dots    &  -9:  & 3 & -18.83(50) & -18.91(50) & \dots      \\  
1973 N                  &  5 & 3.656(54) & 14.55(40) & 14.66(40) & \dots     & 0.28 & \dots    &  +5:  & 3 & -19.85(48) & -19.74(48) & \dots      \\ 
1975 O                  &  3 & 3.701(49) & 15.09(40) & 14.98(40) & \dots     & 0.19 & \dots    &   0:  & 3 & -19.53(47) & -19.64(47) & \dots      \\  
1976 J                  &  5 & 3.637(56) & 14.18(30) & 14.19(30) & \dots     & 0.09 & 0.90(..) &  -2:  & 3 & -20.12(41) & -20.11(41) & \dots      \\  
1980 N                  &  1 & 3.158(79) & 12.40(..) & 12.37(..) & 12.65(..) & 0.07 & 1.28(04) & \dots & 4 & -19.49(50) & -19.52(50) & -19.24(50) \\ 
1981 D                  &  1 & 3.158(79) & 12.50(..) & 12.33(..) & \dots     & 0.07 & \dots    & -15.5 & 5 & -19.39(50) & -19.56(50) & \dots      \\ 
1984 A                  &  1 & 3.072(41) & 12.36(25) & 12.20(25) & \dots     & 0.11 & 1.20(..) & \dots & 4 &  ---       & ---        & \dots      \\ 
1990 N                  &  3 & 3.072(41) & 12.64(03) & 12.62(02) & 12.89(02) & 0.09 & 1.05(05) & -14.0 & 6,8& ---*      & ---*       & ---*       \\ 
1990 O   {\tiny$\surd$} &  1 & 3.958(28) & 16.19(10) & 16.22(08) & 16.65(09) & 0.31 & 0.96(10) &   0.0 & 4 & -19.73(17) & -19.70(16) & -19.27(17) \\ 
1990 T   {\tiny$\surd$} &  1 & 4.080(21) & 17.04(21) & 17.03(16) & 17.31(15) & 0.18 & 1.15(10) & +16.0 & 4 & -19.49(23) & -19.50(19) & -19.22(18) \\
1990 af  {\tiny$\surd$} & -1 & 4.178(17) & 17.77(07) & 17.75(06) & \dots     & 0.12 & 1.56(05) &  -3.0 & 4 & -19.26(11) & -19.28(10) & \dots      \\ 
1991 S\, {\tiny$\surd$} &  3 & 4.222(15) & 17.68(21) & 17.69(16) & 18.02(15) & 0.09 & 1.04(10) & +13.0 & 4 & -19.57(22) & -19.56(18) & -19.23(17) \\ 
1991 T                  &  3 & 3.072(41) & 11.70(06) & 11.51(05) & 11.67(05) & 0.07 & 0.95(05) & -12.0 & 8 &  ---       &  ---       &  ---       \\ 
1991 U   {\tiny$\surd$} &  4 & 3.992(26) & 16.41(21) & 16.35(16) & 16.54(15) & 0.21 & 1.06(10) & +11.0 & 4 & -19.68(25) & -19.74(21) & -19.55(20) \\ 
1991 ag  {\tiny$\surd$} &  3 & 3.617(59) & 14.41(14) & 14.39(15) & 14.72(19) & 0.21 & 0.87(10) &  +7.0 & 4 & -19.79(33) & -19.81(33) & -19.48(35) \\ 
1992 A\, {\tiny$\surd$} &  0 & 3.158(79) & 12.50(07) & 12.50(06) & 12.77(06) & 0.06 & 1.47(05) &  -8.0 & 4a& -19.40(40) & -19.40(40) & -19.13(40) \\ 
1992 J\, {\tiny$\surd$} & -2 & 4.137(19) & 17.64(21) & 17.54(16) & 17.83(15) & 0.19 & 1.56(10) & +14.0 & 4 & -19.18(23) & -19.28(19) & -18.99(18) \\ 
1992 P\, {\tiny$\surd$} &  1 & 3.897(32) & 16.05(07) & 16.09(06) & 16.38(06) & 0.08 & 0.87(10) &  -1.0 & 4 & -19.56(17) & -19.52(17) & -19.23(17) \\ 
1992 ae  {\tiny$\surd$} & -3 & 4.351(11) & 18.50(12) & 18.45(08) & \dots     & 0.12 & 1.28(10) &  +1.0 & 4 & -19.40(13) & -19.45(10) & \dots      \\ 
1992 ag                 &  5 & 3.891(32) & 16.23(08) & 16.15(07) & 16.34(06) & 0.32 & 1.19(10) &  -1.0 & 4 & -19.35(18) & -19.43(17) & -19.24(17) \\ 
1992 al {\tiny$\surd$}  &  3 & 3.627(58) & 14.45(07) & 14.54(06) & 14.88(06) & 0.11 & 1.11(05) &  -5.0 & 4 & -19.80(30) & -19.71(30) & -19.37(30) \\ 
1992 aq {\tiny$\surd$}  &  1 & 4.481(08) & 19.37(09) & 19.33(07) & 19.71(09) & 0.04 & 1.46(10) &   0.0 & 4 & -19.20(10) & -19.24(08) & -18.86(10) \\ 
1992 au {\tiny$\surd$}  & -3 & 4.261(14) & 18.12(21) & 18.11(16) & 18.58(15):& 0.06 & 1.49(10) & +12.0 & 4 & -19.33(22) & -19.34(17) & -18.87(17):\\ 
1992 bc {\tiny$\surd$}  &  2 & 3.773(42) & 15.07(07) & 15.17(06) & 15.54(05) & 0.07 & 0.87(05) & -11.0 & 4 & -19.91(22) & -19.81(22) & -19.44(22) \\
1992 bg {\tiny$\surd$}  &  1 & 4.029(24) & 16.60(08) & 16.66(07) & 16.99(06) & 0.61 & 1.15(10) &  +5.0 & 4 & -19.67(14) & -19.61(14) & -19.28(13) \\ 
1992 bh {\tiny$\surd$}  &  4 & 4.131(19) & 17.59(08) & 17.55(06) & 17.77(06) & 0.07 & 1.05(10) &  -1.0 & 4 & -19.20(12) & -19.24(11) & -19.02(11) \\ 
1992 bk {\tiny$\surd$ } & -3 & 4.240(15) & 18.02(10) & 18.06(07) & 18.27(06) & 0.05 & 1.57(10) &  +8.0 & 4 & -19.32(13) & -19.28(10) & -19.07(10) \\ 
1992 bl {\tiny$\surd$}  &  0 & 4.110(20) & 17.30(08) & 17.33(07) & 17.63(06) & 0.04 & 1.51(10) &  +2.0 & 4 & -19.38(13) & -19.35(12) & -19.05(12) \\ 
1992 bo {\tiny$\surd$}  & -2 & 3.736(45) & 15.74(07) & 15.76(06) & 15.92(05) & 0.09 & 1.69(05) &  -8.0 & 4 & -19.06(24) & -19.04(23) & -18.88(23) \\ 
1992 bp {\tiny$\surd$}  & -2 & 4.373(11) & 18.25(07) & 18.37(06) & 18.70(06) & 0.23 & 1.32(10) &  -2.0 & 4 & -19.77(09) & -19.65(08) & -19.32(08) \\ 
1992 br {\tiny$\surd$}  & -3 & 4.421(10) & 19.24(17) & 19.24(10) & \dots     & 0.09 & 1.69(10) &  +5.0 & 4 & -19.02(18) & -19.02(11) & \dots      \\
1992 bs {\tiny$\surd$}  &  3 & 4.279(13) & 18.30(09) & 18.28(07) & \dots     & 0.04 & 1.13(10) &  +2.0 & 4 & -19.24(11) & -19.26(10) & \dots      \\ 
1993 B                  &  3 & 4.326(12) & 18.48(11) & 18.39(09) & 18.68(10) & 0.26 & 1.04(10) &  +3.0 & 4 & -19.30(13) & -19.39(11) & -19.10(12) \\ 
1993 O  {\tiny$\surd$}  & -2 & 4.193(16) & 17.57(07) & 17.69(06) & 17.96(06) & 0.18 & 1.22(05) &  -6.0 & 4 & -19.53(11) & -19.41(10) & -19.14(10) \\ 
1993 ac {\tiny$\surd$}  & -3 & 4.169(17) & 17.72(16) & 17.71(12) & 17.79(11) & 0.54 & 1.22(10) &  +8.1 & 7 & -19.26(18) & -19.27(15) & -19.19(14) \\
1993 ae {\tiny$\surd$}  & -3 & 3.732(46) & 15.25(20) & 15.31(16) & 15.57(14) & 0.13 & 1.41(10) & +12.8 & 7 & -19.53(30) & -19.47(28) & -19.21(27) \\
1993 ag {\tiny$\surd$}  & -2 & 4.176(17) & 17.83(08) & 17.78(06) & 18.07(06) & 0.37 & 1.32(10) &  -2.0 & 4 & -19.19(12) & -19.24(10) & -18.95(10) \\ 
1993 ah {\tiny$\surd$}  & -1 & 3.935(29) & 16.24(21) & 16.31(16) & 16.65(15) & 0.07 & 1.30(10) & +11.0 & 4 & -19.56(26) & -19.49(22) & -19.15(21) \\ 
1994 D                  & -1 & 3.072(41) & 11.77(07) & 11.80(06) & 12.03(05) & 0.07 & 1.27(10) &  -8.0 & 4a,9& ---      &  ---       &  ---       \\ 
1994 M  {\tiny$\surd$}  & -3 & 3.860(35) & 16.26(08) & 16.24(07) & 16.35(06) & 0.08 & 1.44(10) &  +0.5 & 7 & -19.16(19) & -19.18(19) & -19.07(19) \\
1994 Q  {\tiny$\surd$}  & -1 & 3.940(29) & 16.31(19) & 16.31(20) & 16.60(14) & 0.06 & 1.04(10) & +12.2 & 7 & -19.51(24) & -19.51(25) & -19.22(20) \\ 
1994 S\, {\tiny$\surd$} &  4 & 3.685(51) & 14.78(07) & 14.77(06) & 15.10(06) & 0.07 & 0.92(10) &  -3.6 & 7 & -19.76(26) & -19.77(26) & -19.44(26) \\ 
1994 T                  &  1 & 4.030(24) & 17.24(09) & 17.15(08) & 17.32(07) & 0.10 & 1.40(10) &  +2.2 & 7 & -19.04(15) & -19.13(14) & -18.96(14) \\
1994 ae                 &  5 & 3.201(79) & 13.07(07) & 12.99(06) & 13.34(05) & 0.10 & 0.99(05) & -12.1 & 7 & -19.05(40) & -19.13(40) & -18.78(40) \\ 
1995 D  {\tiny$\surd$}  &  0 & 3.325(79) & 13.18(07) & 13.22(06) & 13.58(06) & 0.19 & 1.00(05) &  -2.6 & 7,10&-19.56(40)& -19.52(40) & -19.16(40) \\
1995 ac                 & 1: & 4.166(17) & 17.08(07) & 17.10(06) & 17.28(06) & 0.14 & 1.01(05) &  -4.4 & 7 & -19.89(11) & -19.87(10) & -19.69(10) \\
1995 ak                 & 1: & 3.824(37) & 16.15(10) & 16.07(09) & 16.13(08) & 0.13 & 1.27(10) &  +3.7 & 7 & -19.09(21) & -19.17(21) & -19.11(20) \\
1995 al                 & 1: & 3.265(79) & 13.31(07) & 13.22(06) & 13.47(06) & 0.05 & 0.94(05) &  -3.9 & 7 & -19.13(40) & -19.22(40) & -18.97(40) \\
1996 C  {\tiny$\surd$}  &  1 & 3.956(28) & 16.54(10) & 16.48(10) & 16.74(08) & 0.05 & 0.92(10) &  +4.0 & 7 & -19.37(17) & -19.43(17) & -19.17(16) \\
1996 X  {\tiny$\surd$}  & -3 & 3.327(79) & 12.97(07) & 13.00(07) & 13.25(07) & 0.23 & 1.29(05) &  -1.7 & 7 & -19.78(40) & -19.75(40) & -19.50(40) \\
1996 ab {\tiny$\surd$}  & 1: & 4.572(07) & 19.52(08) & 19.55(08) & \dots     & 0.11 & 1.16(10) &  +3.3 & 7 & -19.51(09) & -19.48(09) & \dots      \\
1996 bl {\tiny$\surd$}  &  5 & 4.019(24) & 16.67(07) & 16.63(06) & 16.85(07) & 0.31 & 1.15(10) &  -2.2 & 7 & -19.55(14) & -19.59(13) & -19.37(14) \\
1996 bv         & 1: & 3.701(49) & 15.33(13) & 15.19(10) & 15.24(09) & 0.35 & 0.93(10) &  +5.2 & 7 & -19.29(28) & -19.43(26) & -19.38(26) \\
\tablerefs{\footnotesize
(1) Saha et al. 1996b; (2) Leibundgut et al. 1991b; (3) Patat et
al. 1997; (4) Hamuy et al. 1996b; (4a) Hamuy et al. 1996a; (5) Hamuy
et al. 1991; (6) Leibundgut et al. 1991a; (7) Riess et al. 1999,
light-curve template-fitting by us (as prescribed in Hamuy et
al. 1996b, 1996c); (8) Lira et al. 1998; (9) Wu et al. 1995; Richmond
et al. 1995; Patat et al. 1996;  (10) Sadakane et al. 1996.}  
\tablecomments{\footnotesize
An acceptance sign $\surd$ behind the SN name marks a SN that is belonging to the fiducial
sample. An asterisk * indicates a calibrator whose Cepheid-based
absolute magnitude is given in Table 3.}
\enddata
\end{deluxetable}

\clearpage

\begin{deluxetable}{llcccccccc}
\tablenum{2}
\tabletypesize{\tiny}
\tablewidth{0pt}
\tablecaption{Host Galaxy and Positional Parameters of Blue SNe Ia\label{tbl2}}
\tablehead{
\colhead{SN}       & \colhead{Galaxy}   & \colhead{Type}     & \colhead{T}        &
\colhead{$\alpha$} & \colhead{$\delta$} & \colhead{D$_{25}$} & \colhead{E/W}      &
\colhead{N/S}      & \colhead{r$_{\rm offset}$/r$_{25}$} \\ 
\colhead{(1)} & \colhead{(2)} & (3) & (4) & (5) & (6) & (7) & (8) & (9) & (10) \\}
\startdata
1895B   & NGC 5253     & Am   & 5: & 133956 & -313841 & 301 & +16 & +23 & 0.19 \\ 
1937 C  & IC 4182      & Im   & 5  & 130545 & +373621 & 362 & +30 & +40 & 0.28 \\ 
1956 A  & NGC 3992     & Sb   & 3  & 115736 & +532231 & 455 & +67 &  -9 & 0.30 \\ 
1959 C  & UGC 8263     & Sc   & 5  & 131123 & +032442 &  81 & +7  &  -3 & 0.19 \\ 
1960 F  & NGC 4496A    & Sc   & 5  & 123140 & +035621 & 238 & +38 & +24 & 0.38 \\ 
1962 A  & MCG 5-31-132 & E/S0 & -2 & 130636 & +275224 &  49 & -11 &  +7 & 0.53 \\ 
1965 I  & NGC 4753     & S0p  & -1 & 125223 & -011157 & 362 & -98 & +68 & 0.66 \\ 
1966 K  & UGC 6322     & S0p: & -1 & 111812 & +281600 &  83 & -26 & +15 & 0.72 \\ 
1967 C  & NGC 3389     & Sc   & 5  & 104828 & +123201 & 165 & -43 & +44 & 0.75 \\ 
1969 C  & NGC 3811     & Sc   & 5  & 114116 & +474135 & 131 & +9  &  +6 & 0.17 \\ 
1970 J  & NGC 7619     & E    & -3 & 232015 & +081223 & 151 & -27 & -30 & 0.53 \\ 
1971 G  & NGC 4165     & Sa   & 1  & 121212 & +131448 &  72 & +3  & -30 & 0.84 \\ 
1971 L  & NGC 6384     & Sb   & 3  & 173225 & +070337 & 370 & +27 & +20 & 0.18 \\ 
1972 E  & NGC 5253     & Am   & 5: & 133956 & -313841 & 301 & -38 &-100 & 0.71 \\ 
1972 H  & NCG 3147     & Sb   & 3  & 101653 & +732404 & 256 & +31 & +37 & 0.38 \\ 
1972 J  & NGC 7634     & S0   & -1 & 232142 & +085314 &  74 & -5  & -30 & 0.82 \\ 
1973 N  & NGC 7495     & Sc   & 5  & 230854 & +120200 & 109 & -14 &  -7 & 0.29 \\ 
1974 G  & NGC 4414     & Sc   & 5  & 122627 & +311329 & 228 & +27 & -56 & 0.54 \\ 
1975 O  & NGC 2487     & Sb   & 3  & 075820 & +250859 & 154 & +26 & +15 & 0.39 \\ 
1976 J  & NGC 977      & S:   & 5  & 023304 & -104536 & 117 & -10 & -25 & 0.73 \\ 
1980 N  & NGC 1316     & Sap  & 1  & 032242 & -371227 & 721 & +220& -20 & 0.61 \\ 
1981 B  & NGC 4536     & Sbc  & 4  & 123427 & +021119 & 455 & +41 & +41 & 0.25 \\ 
1981 D  & NGC 1316     & Sap  & 1  & 032242 & -371227 & 721 & -20 &-100 & 0.28 \\ 
1984 A  & NGC 4419     & Sa   & 1  & 122657 & +150252 & 199 & -15 & +30 & 0.34 \\ 
1989 B  & NGC 3627     & Sb   & 3  & 112014 & +125942 & 547 & -15 & +50 & 0.19 \\ 
1990 N  & NGC 4639     & Sb   & 3  & 124253 & +131531 & 165 & +63 &  -2 & 0.76 \\
1990 O  & MCG 3-44-03  & Sa   & 1  & 171533 & +161842 &  52 & +20 &  -3 & 0.78 \\ 
1990 T  & PGC 63925    & Sa   & 1  & 195900 & -561530 &  81 & +25 &  -2 & 0.62 \\ 
1990 af & Anon 2135-62 & S0   & -1 & 213500 & -624400 &\dots&  -7 &  +7 &\dots \\ 
1991 S  & UGC 5691     & Sb   & 3  & 102932 & +215937 &  64 &  +4 & +17 & 0.55 \\ 
1991 T  & NGC 4527     & Sb   & 3  & 123409 & +023913 & 370 & +26 & +45 & 0.28 \\ 
1991 U  & IC 4232      & Sbc  & 4  & 132322 & -260639 &  67 &  -3 &  +6 & 0.20 \\ 
1991 ag & IC 4919      & Sb   & 3  & 200009 & -552228 &  89 &  -3 & +22 & 0.50 \\ 
1992 A  & NGC 1380     & S0/a & 0  & 033627 & -345833 & 287 &  -3 & +62 & 0.43 \\ 
1992 J  & Anon 1009-26 & E/S0 & -2 & 100900 & -263900 &\dots& -11 & +13 &\dots \\ 
1992 P  & IC 3690      & Sa   & 1  & 124250 & +102134 &  66 &  -6 & +11 & 0.38 \\ 
1992 ae & Anon 2128-61 & E    & -3 & 212818 & -613300 &\dots&  +3 &  +5 &\dots \\ 
1992 ag & ESO 508-G67  & S:   & 5  & 132410 & -235243 &  61 &  -4 &   0 & 0.13 \\ 
1992 al & ESO 234-G69  & Sb   & 3  & 204554 & -512332 & 125 & +18 & -12 & 0.35 \\ 
1992 aq & Anon 2304-37 & Sa   & 1  & 230436 & -372100 &\dots&  +2 &  -7 &\dots \\ 
1992 au & Anon 0010-49 & E    & -3 & 001036 & -49560  &\dots& +21 &  +9 &\dots \\ 
1992 bc & ESO 300-G9   & Sab  & 2  & 030516 & -393337 &  59 & +15 &  +5 & 0.54 \\ 
1992 bg & Anon 0741-62 & Sa   & 1  & 074154 & -623100 &\dots&  -3 &  +6 &\dots \\ 
1992 bh & Anon 0459-58 & Sbc  & 4  & 045930 & -585000 &\dots&  +1 &-2.5 &\dots \\ 
1992 bk & ESO 156-G8   & E    & -3 & 034301 & -533815 &  79 & +12 & +21 & 0.61 \\ 
1992 bl & ESO 291-G11  & S0/a & 0  & 231512 & -444414 &  43 & +15 & -22 & 1.23 \\ 
1992 bo & ESO 352-G57  & E/S0 & -2 & 012202 & -341150 &  81 & -47 & -55 & 1.79 \\ 
1992 bp & Anon 0336-18 & E/S0 & -2 & 033636 & -182100 &\dots&  -6 &-1.5 &\dots \\ 
1992 br & Anon 0145-56 & E0   & -3 & 014542 & -560500 &\dots&  +3 &  -7 &\dots \\ 
1992 bs & Anon 0329-37 & Sb   & 3  & 032930 & -371600 &\dots&  -9 &  +4 &\dots \\ 
1993 B  & Anon 1034-34 & Sb   & 3  & 103454 & -342700 &\dots&  +1 &  +5 &\dots \\ 
1993 O  & Anon 1331-33 & E/S0 & -2 & 133106 & -331200 &\dots& -14 &  +8 &\dots \\ 
1993 ac & PGC 17787    & E    & -3 & 054616 & +632110 &\dots&  -5 & +31 &\dots \\
1993 ae & UGC 1071     & E    & -3 & 012945 & -015831 &  83 & +16 & +23 & 0.68 \\
1993 ag & Anon 1003-35 & E/S0 & -2 & 100336 & -352800 &\dots&  -5 &  -6 &\dots \\
1993 ah & ESO 471-G27  & S0   & -1 & 235151 & -275748 &  60 &  -1 &  +8 & 0.27 \\ 
1994 D  & NGC 4526     & S0   & -1 & 123403 & +074201 & 435 &  -9 &  +7 & 0.05 \\ 
1994 M  & NGC 4493     & E    & -3 & 123108 & +003648 &  74 &  +3 & -28 & 0.76 \\ 
1994 Q  & PGC 59076    & S0   & -1 & 164951 & +402559 &  32 &  -1 &  -4 & 0.26 \\ 
1994 S  & NGC 4495     & Sbc  & 4  & 123123 & +290813 &  83 & -13 &  -7 & 0.36 \\ 
1994 T  & PGC 46640    & Sa   & 1  & 132129 & -020941 &\dots&  +4 & -12 &\dots \\     
1994 ae & NGC 3370     & Sc   & 5  & 104704 & +171626 & 190 & -30 &  +6 & 0.32 \\ 
1995 D  & NGC 2962     & S0/a & 0  & 094054 & +051000 & 158 & +11 & -91 & 1.16 \\
1995 ac & Anon 2245-08 & S:   & 1  & 224542 & -084500 &\dots&  -1 &  -1 &\dots \\ 
1995 ak & IC 1844      & S:   & 1  & 024549 & +031348 &  47 &  -7 &  +1 & 0.30 \\
1995 al & NGC 3021     & Sbc  & 4  & 095057 & +333316 &  87 & -15 &  -3 & 0.35 \\
1996 C  & MCG +08-25-47& Sa   & 1  & 135048 & +492000 &  66 &  -2 & +13 & 0.40 \\
1996 X  & NGC 5061     & E0   & -3 & 131805 & -265010 & 213 & -51 & -32 & 0.57 \\
1996 ab & Anon 1521+27 & S:   & 1  & 152106 & +275500 &\dots& +2: & +1: &\dots \\ 
1996 bl & Anon 0036+11 & Sc   & 5  & 003617 & +112340 &\dots&  -3 &  +6 &\dots \\
1996 bv & UGC 3432     & Scd: & 5  & 061612 & +570200 & 100 &  -2 &  +2 & 0.06 \\
1998 bu & NGC 3368     & Sab  & 2  & 104645 & +114917 & 456 &  +4 & +55 & 0.24 \\
\enddata
\end{deluxetable}

\clearpage

\begin{table}
\footnotesize
\tablenum{3}
\begin{center}
\caption{\sc Absolute B, V, and I magnitudes of blue SNe\,Ia calibrated through Cepheid distances of their parent galaxies \label{tbl3}}
\begin{tabular}{llllllcrrrc}
\tableline \tableline 
SN & Galaxy & $\log v$\tablenotemark{a} & (m-M)$_{AB}$ & (m-M)$_{AV}$ & (m-M)$^0$ & ref. & B$_{AB}$ &  V$_{AV}$ &  I$_{AV}$ & ref. \\
(1) & (2) & (3) & (4) & (5) & (6) & (7) & (8) & (9) & (10) & (11) \\ 
\tableline
1895 B & NGC 5253 & 2.464\tablenotemark{b}& 28.13(08) & 28.10(07) & \nodata   & 2 &  8.26(20) & \nodata   &\nodata    &  9   \\
1937 C & IC 4182  & 2.519                 & 28.36(09) & 28.36(12) & \nodata   & 1 &  8.80(09) &  8.82(11) &\nodata    & 10   \\
1960 F & NGC 4496A& 3.072\tablenotemark{c}& 31.16(10) & 31.13(10) & \nodata   & 3 & 11.60(15) & 11.51(20) &\nodata    &  3   \\
1972 E & NGC 5253 & 2.464\tablenotemark{b}& 28.13(08) & 28.10(07) & \nodata   & 2 &  8.49(14) &  8.49(15) &  8.80(19) & 10   \\
1974 G & NGC 4414 & 2.820                 & \nodata   & \nodata   & 31.46(17) & 4 & 12.48(05) & 12.30(05) &\nodata    & 11   \\
1981 B & NGC 4536 & 3.072\tablenotemark{c}& \nodata   & \nodata   & 31.10(12) & 5 & 12.03(03) & 11.93(03) &\nodata    & 10   \\
1989 B & NGC 3627 & 2.734                 & \nodata   & \nodata   & 30.22(12) & 6 & 12.34(05) & 12.02(05) & 11.75(05) & 12   \\
1990 N & NGC 4639 & 3.072\tablenotemark{c}& \nodata   & \nodata   & 32.03(22) & 7 & 12.75(03) & 12.71(02) & 12.94(02) & 10,13\\
1998 bu& NGC 3368 & 2.814\tablenotemark{d}& \nodata   & \nodata   & 30.37(16) & 8 & 12.18(03) & 11.88(03) & 11.67(05) & 14   \\
\tableline
\end{tabular}
\begin{tabular}{llcrrrcccr}
\\
\tableline
SN & E$_{B-V}$ & ref. & B$^0$ & V$^0$ & I$^0$ & M$_B^0$ & M$_V^0$ & M$_I^0$ & $\Delta$m$_{15}$\\
(1) & (12) & (13) & (14) & (15) & (16) & (17) & (18) & (19) & (20) \\ 
\tableline
1895 B &\nodata    &\nodata&\nodata    &\nodata    &\nodata    &-19.87(22) &\nodata    &\nodata    &\nodata   \\ 
1937 C &\nodata    &\nodata&\nodata    &\nodata    &\nodata    &-19.56(15) &-19.54(17) &\nodata    & 0.87(10) \\ 
1960 F &\nodata    &\nodata&\nodata    &\nodata    &\nodata    &-19.56(18) &-19.62(22) &\nodata    & 1.06(12) \\ 
1972 E &\nodata    &\nodata&\nodata    &\nodata    &\nodata    &-19.64(16) &-19.61(17) &-19.27(20) & 0.87(10) \\ 
1974 G &  0.16(07) &    11 & 11.79(31) & 11.77(24) &\nodata    &-19.67(34) &-19.69(27) &\nodata    & 1.11(06) \\ 
1981 B &  0.10(03) &     5 & 11.60(13) & 11.60(10) &\nodata    &-19.50(18) &-19.50(16) &\nodata    & 1.10(07) \\
1989 B &  0.37(03) &    12 & 10.75(14) & 10.80(11) & 11.01(08) &-19.47(18) &-19.42(16) &-19.21(14) & 1.31(07) \\ 
1990 N &  0.026(03)&    15 & 12.64(13) & 12.62(10) & 12.89(09) &-19.39(26) &-19.41(24) &-19.14(23) & 1.05(05) \\ 
1998 bu&  0.365(06)&    14 & 10.61(26) & 10.68(20) & 10.94(13) &-19.76(31) &-19.69(26) &-19.43(21) & 1.08(05) \\ 
\tableline
\multicolumn{6}{r}{mean (straight,$\;\;$ excl. SN 1895 B)} & -19.57(04)& -19.56(04) & -19.26(06) & 1.06(05) \\
\multicolumn{6}{r}{mean (weighted, excl. SN 1895 B)} & -19.55(07)& -19.53(06) & -19.25(09) & 1.08(02)
\end{tabular}
\end{center}
\tablenotetext{a}{\scriptsize The velocities used are corrected for Virgocentric infall assuming a local infall velocity of 220 km\,s$^{-1}$}
\tablenotetext{b}{\scriptsize The mean velocity v=291 km\,s$^{-1}$ of the Cen A group is used}
\tablenotetext{c}{\scriptsize The mean velocity v=1179 km\,s$^{-1}$ of the Virgo group is used}
\tablenotetext{d}{\scriptsize The mean velocity v=652 km\,s$^{-1}$ of the Leo group is used}
\tablerefs{\scriptsize (1) Saha et al. 1994; (2) Saha et al. 1995; (3) Saha et al. 1996b; (4) Turner et al. 1998 ; (5) Saha et al. 1996a; 
(6) Saha et al. 1999; (7) Saha et al. 1997; (8) Tanvir et al. 1995; (9) Schaefer 1995; (10) Hamuy et al. 1996a; (11) Schaefer 1998; 
(12) Wells et al. 1994; (13) Lira et al. 1998; (14) Suntzeff et al. 1999; Jha et al. 1999; (15) Schlegel, Finkbeiner, \& Davies 1998.} 
\end{table}

\clearpage

\begin{table}
\tablenum{4}
\caption{\sc Intrinsic colors of unreddened blue SNe Ia after 1985 \label{tbl4}}
\begin{tabular}{lcccc}
\\
\tableline\tableline
 & (B-V) & n & (V-I) & n \\
\tableline
SNe\,Ia in E/S0s                          & -0.013(015) & 16 & -0.240(024) & 12 \\
SNe\,Ia in spirals with $r/r_{25}\ge0.4$\\
\hspace{85pt} and (B-V)$\le$0.06          & -0.013(015) &  9 & -0.326(018) &  9 \\
Calibrators                               & -0.009(015) &  8 & -0.270(027) &  4 \\
mean                                      & -0.012(008) & 33 & -0.276(016) & 25 \\
other SNe\,Ia in spirals                  & -0.001(016) & 10 & -0.288(024) &  8 \\
\tableline
\end{tabular}
\end{table}
 
\clearpage

\begin{table}
\tablenum{5}
\caption{\sc Reddened SNe Ia of the present sample \label{tbl5}}
\begin{tabular}{lcccccccccc}
\\
\tableline\tableline
SN & r/r$_{25}$ & (B-V) & (V-I) & $<$E(B-V)$>$ & B$^0$ & V$^0$ & I$^0$ & M$_B^{60}$ & M$_V^{60}$ & M$_I^{60}$\\
(1) & (2) & (3) & (4) & (5) & (6) & (7) & (8) & (9) & (10) & (11)\\
\tableline
1992 ag & 0.13   & 0.08 & -0.19 & 0.079 & 15.89 & 15.89 & 16.18 & -19.67 & -19.67 & -19.38 \\  
1993 B  &\nodata & 0.09 & -0.29 & 0.067 & 18.19 & 18.17 & 18.55 & -19.55 & -19.57 & -19.19 \\ 
1994 T  &\nodata & 0.09 & -0.17 & 0.090 & 16.85 & 16.85 & 17.14 & -19.41 & -19.41 & -19.12 \\ 
1994 ae & 0.32   & 0.08 & -0.35 & 0.049 & 12.86 & 12.83 & 13.24 & -19.25 & -19.28 & -18.87 \\ 
1995 ak & 0.30   & 0.08 & -0.06 & 0.105 & 15.70 & 15.72 & 15.92 & -19.53 & -19.51 & -19.31 \\
1995 al & 0.35   & 0.09 & -0.25 & 0.074 & 12.99 & 12.98 & 13.32 & -19.44 & -19.45 & -19.11 \\
1996 bv & 0.06   & 0.14 & -0.05 & 0.147 & 14.70 & 14.70 & 14.95 & -19.91 & -19.91 & -19.66 \\
\tableline
\end{tabular}
\end{table}

\end{document}